\documentclass{elsart}

    \usepackage{latexsym,bm,amsmath,amssymb,amsfonts}
    \usepackage{epsfig,graphics,graphicx}
    \usepackage{makeidx}
    \usepackage{citesort}
    \usepackage{slashed}
    \usepackage{float,multicol}

\newcommand{\cO}{{\cal O}}
\newcommand{\avg}[1]{\left\langle #1 \right\rangle}

    \newcommand{\dif}{{\rm d}}
    \newcommand{\abar}{\bar{\alpha}_s}
    \newcommand{\atpi}{\frac{\bar{\alpha}_s}{2\pi}}
    \newcommand{\del}{\partial}
    \newcommand{\lan}{\left\langle}
    \newcommand{\ran}{\right\rangle}
    \newcommand{\cal}{\mathcal}

    \newcommand{\grad}{\nabla}
    
    \newcommand{\lap}[1]{\nabla_{\bm{#1}}^2}
    \newcommand{\nn}{\nonumber\\}

    \long\def\comment#1{ }
    
    \newcommand{\beq}{\vspace{-.4cm}\begin{eqnarray}}
    \newcommand{\eeq}{\vspace{-.5cm}\end{eqnarray}}

\def\simge{\mathrel{%
   \rlap{\raise 0.511ex \hbox{$>$}}{\lower 0.511ex \hbox{$\sim$}}}}
\def\simle{\mathrel{
   \rlap{\raise 0.511ex \hbox{$<$}}{\lower 0.511ex \hbox{$\sim$}}}}

\begin{document}

\begin{flushright}
~\vspace{-1.25cm}\\
{\small\sf SACLAY--T05/138\\ECT$^*$--05--12}
\end{flushright}
\vspace{2.cm}

\begin{frontmatter}

\parbox[]{16.0cm}{ \begin{center}
\title{On the Probabilistic Interpretation of the Evolution
Equations with Pomeron Loops in QCD}

\author{E.~Iancu\thanksref{th2}},
\author{G.~Soyez\thanksref{th3}}
\author{and D.N.~Triantafyllopoulos\thanksref{th4}}

\address{Service de Physique Th\'eorique, CEA/DSM/SPhT,  Unit\'e de recherche
associ\'ee au CNRS (URA D2306), CE Saclay,
        F-91191 Gif-sur-Yvette, France}

\thanks[th2]{Membre du Centre National de la Recherche Scientifique
(CNRS), France.}
\thanks[th3]{On leave from the Fundamental Theoretical Physics group
of the University of Li\`ege.}
\thanks[th4]{Present address:  ECT$^*$, Villa Tambosi, Strada delle Tabarelle
286, I-38050 Villazzano (TN), Italy}
\date{\today}
\vspace{0.8cm}
\begin{abstract}

We study some structural aspects of the evolution equations with
Pomeron loops recently derived in QCD at high energy and for a
large number of colors, with the purpose of clarifying their
probabilistic interpretation.  We show that, in spite of their
appealing dipolar structure and of the self--duality of the
underlying Hamiltonian, these equations cannot be given a
meaningful interpretation in terms of a system of dipoles which
evolves through dissociation (one dipole splitting into two) and
recombination (two dipoles merging into one). The problem comes
from the saturation effects, which cannot be described as dipole
recombination, not even effectively. We establish this by showing
that a (probabilistically meaningful) dipolar evolution in either
the target or the projectile wavefunction cannot reproduce the
actual evolution equations in QCD.

\end{abstract}
\end{center}}

\end{frontmatter}
\newpage

\section{Introduction}
\setcounter{equation}{0}\label{SECT_INTRO}

Over the last year, an intense activity in the field of
high--energy QCD has been triggered by the observations that
\texttt{(i)} the gluon number fluctuations at low energy play an
important role in the evolution towards gluon saturation with
increasing energy \cite{IM032,MS04,IMM04}, and \texttt{(ii)} the
relevant fluctuations are in fact missed \cite{IT04} by the
existing approaches to non--linear evolution in QCD at high
energy, namely, the Balitsky--JIMWLK equations
\cite{B,JKLW,RGE,W}. To cope with that, new equations have been
proposed \cite{IT04,IT05,MSW05,BIIT05} (see also Refs.
\cite{LL05,Levin05}), which include the particle number
fluctuations together with the saturation effects in the limit
where the number of colors $N_c$ is large. At the same time, an
ambitious program was launched, which aims at generalizing these
equations to arbitrary values of $N_c$
\cite{KL05,KL3,KL4,BREM,MMSW05,HIMS05,Balit05,KL5}. This effort
led already to some important results  --- in particular, the
recognition \cite{KL3,BIIT05} of a powerful self--duality property
of the high--energy evolution, and the construction of an
effective Hamiltonian which is explicitly self--dual
\cite{BREM,Balit05}
---, but the general problem is still under study, and the
evolution equations for arbitrary $N_c$ are not yet known.

In this paper, we shall restrict ourselves to the large--$N_c$
limit and investigate whether the evolution equations proposed in
Refs. \cite{IT04,IT05} can be given a probabilistic interpretation
in the sense of the dipole picture. To motivate our search, let us
remind here that most of the modern approaches to high--energy
evolution in QCD have an underlying probabilistic structure, which
makes their physical interpretation transparent and greatly
simplifies the analysis of their consequences, via analytic or
numerical techniques.

Specifically, the {\em color dipole picture} introduced by Mueller
\cite{AM94,AM95,IM031} describes the BFKL  evolution \cite{BFKL}
at large $N_c$ (including the particle number fluctuations alluded
to above) as a classical Markovian process in which a system of
dipoles evolves through {\em dipole splitting}. This manifest
probabilistic interpretation has enabled Salam to build up a
Monte--Carlo code allowing for numerical studies of the dipole
evolution \cite{Salam95}. The dipole picture is also underlying
the derivation by Kovchegov \cite{K} of the non--linear
Balitsky--Kovchegov (BK) equation. Recently, this derivation has
been extended, by Levin and Lublinsky \cite{LL04}, to the
large--$N_c$ version of the Balitsky--JIMWLK hierarchy.

Similarly, the {\em color glass condensate} formalism
\cite{MV,RGE,CGCreviews}, which aims at a description of the high
density environment at, or near, saturation, is explicitly
formulated in terms of a classical probability distribution, which
evolves with increasing energy according to a functional
Fokker--Planck equation, the JIMWLK equation \cite{JKLW,RGE,W}.
The basic microscopic process at work here (in addition to BFKL
evolution) is {\em gluon recombination} : the JIMWLK Hamiltonian
includes vertices for $n\to 2$ gluon merging with $n> 2$, which
lead to gluon saturation at high energy. The stochastic nature of
the JIMWLK evolution is most clearly exhibited by its
reformulation as a Langevin equation describing a random walk in
the space of Wilson lines \cite{PATH}. This is also the most
convenient formulation
for numerical studies, as demonstrated by the analysis in Ref.
\cite{RW03}.

Note that the color dipole picture can be also formulated as a
``color glass'' \cite{IM031,MSW05}, but the ensuing picture is
complementary to the JIMWK evolution: it describes gluon splitting
(at large--$N_c$), but not also gluon recombination. Clearly, a
complete picture should involve a synthesis of these two
approaches, but such a synthesis has not been yet achieved in a
wavefunction formalism, not even at large $N_c$. (See however the
recent developments in Refs. \cite{BREM,Balit05}.) On the other
hand, the equations constructed in Refs. \cite{IT04,IT05} propose
such a synthesis at the level of the {\em dipole scattering
amplitudes} (and for large $N_c$).

Specifically, these equations refer to the scattering between a
projectile which is a set of dipoles and a generic target. As
shown in Ref. \cite{BIIT05}, the equations admit a natural {\em
Pomeron} interpretation, where the ``Pomeron'' is the scattering
amplitude for an external dipole: They describe the BFKL evolution
of the Pomerons together with their interactions: $1\to 2$ Pomeron
splitting with the standard dipole vertex \cite{AM94}, and $2\to
1$ Pomeron merging with the ``triple--Pomeron vertex''
\cite{BW93,BV99}. By iterating these equations, one generates
Pomeron loops \cite{AM95,IT04,BIIT05}. So, from now on, we shall
refer to them as `the evolution equations with Pomeron loops'.

The general structure of these equations alluded to above makes it
tempting to search for a probabilistic interpretation in terms of
a {\em dipolar reaction--diffusion process}, that is, a classical
stochastic process in which a collection of dipoles --- which
represents either the target or the projectile --- evolves through
splitting and recombination. In what follows we shall briefly
explain (leaving the details to the main body of the paper) why
such an interpretation is tantalizing in the first place, what
would be its benefits in practice, and why it does not really
exist. But before we proceed, let us emphasize here that, although
our main conclusion is essentially negative --- we demonstrate the
{\em non}--existence of a dipole interpretation
---, this result has also positive implications, in the sense that
it helps reorienting the efforts towards solving these equations
(in particular, via numerical methods), and also towards better
understanding the mechanism for gluon saturation. Besides, as we
shall see, it helps clarifying some confusion which exists on this
point in the literature.

The reason why such an interpretation looks at the first sight
appealing is because, taken separately, each of the terms
appearing in these equations describes dipole splitting {\em when
considered in an appropriate Lorentz frame} (and for large $N_c$).
Specifically, the non--linear terms responsible for unitarity
corrections can be seen as dipole splitting in the wavefunction of
the {\em projectile} \cite{B,K,Janik,LL04}, whereas the
fluctuation terms relevant at low energy represent dipole
splitting in the wavefunction of the {\em target}
\cite{AM94,IMM04,IT04} (see Sect. \ref{SECT_TEQS} below for
details). Moreover, the Hamiltonian underlying the Pomeron loop
equations, as constructed in Ref. \cite{BIIT05}, has a remarkable
{\em self--duality property} \cite{KL3,BIIT05}, which in a given
frame interchanges the splitting and the merging pieces of the
Hamiltonian. Since the splitting piece has a well--founded dipolar
interpretation, it is then tempting to search for a similar
interpretation for the merging piece as well.

Indeed, a fully dipolar picture of the evolution of the
(projectile or target) wavefunction, involving dipole splitting
{\em and} merging, would be highly desirable, since very useful in
practice. For instance, this would allow one to extend the
Monte--Carlo method by Salam \cite{Salam95} up to arbitrarily high
energy, and thus facilitate the numerical study of the evolution
equations with Pomeron loops. More generally, this would establish
an explicit correspondence between high--energy evolution in QCD
and some Markovian ``reaction--diffusion processes'' which are
intensively studied in the context of statistical physics (see the
textbook \cite{Gardiner}, and the review papers \cite{SaarPanja}
for more recent developments), from where one could borrow some
methods to QCD. Such a correspondence has been already proven to
be useful, both at the level of the mean field approximation (the
BK equation) \cite{MP03}, where it shed a new light on the
`geometric scaling' property of the solution
\cite{geometric,SCALING,MT02,DT02}, and in analyzing the role of
the particle number fluctuations under asymptotic conditions (very
high energy and arbitrarily small coupling constant)
\cite{IMM04,IT04}. But, clearly, it would be extremely useful to
extend this correspondence to a more interesting, non--asymptotic,
regime, and rely on it in order to translate to QCD the huge
experience accumulated with this problem in the context of
statistical physics. Last but not least, a wavefunction picture
would allow for more general studies, and for the calculation of
quantities other than the dipole scattering amplitudes (so like
the gluon production in onium--onium scattering, a problem
recently addressed in Ref. \cite{Kov05}).

In view of the above, it may seem a little disappointing that such
a dipolar picture cannot be actually given, for the reasons that
we explain now. The problem comes from the {\em gluon saturation}
effects, which cannot be described as dipole recombination. This
problem can be seen already when trying to introduce saturation
effects in the original dipole picture by Mueller
\cite{AM94,AM95}: when two dipoles within the same wavefunction
are allowed to interact with each other by exchanging gluons, they
produce not only dipoles, but also more complicated color states,
like quadrupoles. This illustrates the fact that the merging
processes responsible for saturation involve a non--trivial flow
of color, and thus transcend the dipole picture. The actual
theoretical tool for describing saturation, the JIMWLK equation
\cite{JKLW,RGE,W}, is written in terms of colorful degrees of
freedom (gluons and the associated gauge fields), and cannot be
directly simplified by taking the large--$N_c$ limit
\cite{BIIT05}. Such a simplification becomes possible only after
applying the JIMWLK equation to dipole scattering amplitudes (thus
yielding the Balitsky equations \cite{B}); but in that case, the
ensuing dipole interpretation refers to the evolution of the
projectile through dipole splitting, and not to that of the target
towards gluon saturation.

This being said, the suggestive structure of the evolution
equations with Pomeron loops alluded to above makes it tempting to
search for an {\em effective} dipole picture at least, where by
``effective'' we mean a fictitious dipolar reaction--diffusion
process which reproduces the actual equations for the scattering
amplitudes without correctly reflecting the microscopic processes
at work. (Such an effective picture would be, of course, still
suitable for Monte--Carlo simulations.) And as a matter of facts,
an effective dipole picture of this type has been recently
proposed in Ref. \cite{LL05}, from the point of view of projectile
evolution. However, as we shall demonstrate in what follows, that
picture is truly ill--defined (it does not describe a physically
acceptable reaction--diffusion process), and thus is useless in
practice\footnote{Besides, the analysis in Ref. \cite{LL05}
involves some errors of sign that we shall correct in our
corresponding analysis in Sect. \ref{SECT_PROJ}.}.

The basic problem is that the $2 \to 1$ vertex which is supposed
to describe dipole recombination within this effective picture
(this is the same as the ``triple--Pomeron vertex''
\cite{BW93,BV99}) has {\em no definite sign}, and thus it cannot
be given a probabilistic interpretation, as a merging rate per
unit rapidity. Very likely, this difficulty reflects the fact that
the actual physical mechanism responsible for gluon saturation in
QCD is a {\em coherent}, collective, effect\footnote{This is most
transparently seen within the random walk formulation
\cite{W,PATH} of the JIMWLK evolution, where the probability for
emitting new gluons at each step in the evolution involves Wilson
lines and saturates to a field--independent value in the
high--energy limit.} --- the suppression of the rate for gluon
emission in the high density regime by strong color fields
\cite{W,SAT,AM02,PATH} ---, and not a `mechanical' reduction in
the number of gluons. And indeed, the effective `dipole
recombination' process mentioned above has the property to leave
the total number of dipoles in the system {\em unchanged} (unlike
a real merging mechanism, which always decreases the number of
particles).

The above considerations motivate our strategy in this paper,
which will be to demonstrate the non--existence of the dipolar
picture through explicit construction. Namely, we shall consider a
generic dipole--like reaction--diffusion process with arbitrary,
non--local, vertices for dipole splitting and merging, and show
that the only way for this process to reproduce the actual
equations with Pomeron loops in QCD is by choosing the `dipole
recombination vertex' as the triple--Pomeron vertex mentioned
before. Since the latter has no well--defined probabilistic
interpretation, this choice renders the whole process meaningless.

For our analysis to be convincing, we shall repeat it from the
points of view of both target and projectile evolution. Indeed,
{\em a priori}, these two points of view are not fully equivalent
with each other, as they involve different `factorization schemes'
connecting the scattering amplitudes to the dipole densities:

In the context of target evolution, it will be natural to evaluate
the scattering amplitude for an external dipole {\em in the
two--gluon exchange approximation}. This is motivated by the fact
that \texttt{(i)} the dipole picture applies {\em a priori} to the
dilute target regime, where this approximation is correct indeed,
and  \texttt{(ii)} when extended to the high--density regime, this
approximation leads to a self--dual Hamiltonian  \cite{BIIT05},
which has a suggestive, dipole--like, structure and thus is a
natural starting point for the search of a dipole picture.

From the perspective of projectile evolution, on the other hand,
there is {\em a priori} no need for further approximations (except
for the assumed reaction--diffusion process): Following a strategy
by Kovchegov \cite{K} (see also Refs. \cite{LL03,LL04,LL05}), the
equations obeyed by the dipole densities can be exactly mapped
onto equations for the scattering amplitudes, including multiple
scattering effects to all orders. Yet, the two--gluon exchange
approximation strikes back when one attempts to deduce the vertex
for dipole recombination by comparison with the actual equations
in QCD.

Let us finally note that the lack of a probabilistic
interpretation in terms of dipoles does not {\em a priori}
preclude the existence of alternative stochastic descriptions,
involving some more elementary degrees of freedom, so like the
gauge field $\alpha_a$ to be introduced in Sect.
\ref{SECT_TARGET}. For instance, we have already mentioned that,
within the JIMWLK formalism, the saturation processes have a well
defined probabilistic interpretation \cite{PATH}, but in terms of
gluon fields. Moreover, it is also known that the evolution
equations with Pomeron loops can be compactly summarized into a
single Langevin equation, with a rather unusual noise term though
\cite{IT05,MSW05}; this is again suggestive of a classical
stochastic process. An approximate version of this Langevin
equation \cite{IT04} has been recently used for numerical studies
\cite{GS05,EGBM05}. It would be very interesting to clarify the
precise nature of the stochastic process underlying the equations
with Pomeron loops, if any.

The structure of this paper is as follows: In Sect.
\ref{SECT_TEQS} we briefly review the evolution equations with
Pomeron loops together with their physical interpretation from the
perspectives of both target and projectile evolution. In Sect.
\ref{SECT_TARGET}, we study the possibility to interpret these
equations in terms of dipole evolution in the target. To that aim,
we rely on the Hamiltonian formulation introduced in Ref.
\cite{BIIT05} (to be reviewed in Sect. \ref{SECT_ALPHA}), that we
first rewrite in terms of dipole operators (in Sect.
\ref{SECT_HD}) and then use to deduce evolution equations for the
dipole densities (in Sect. \ref{SECT_TDIP}). In the next section,
these equations are compared to those generated by a genuine
reaction--diffusion process. Specifically, in Sect.
\ref{SEC_MASTER} we introduce the Master equation defining a
generic stochastic process with non--local vertices for splitting
and recombination, and use it to deduce evolution equations for
the dipole densities. (For convenience, the general equations are
summarized in the Appendix.) Finally, in Sect. \ref{SECT_PROJ}, we
consider a reaction--diffusion process taking place at the level
of the projectile. From the equations for the dipole densities
(cf. Sect. \ref{SEC_MASTER}), we deduce equations for the
scattering amplitudes, which are eventually compared to the QCD
equations with Pomeron loops.

\section{Evolution equations with Pomeron loops}
\setcounter{equation}{0}\label{SECT_TEQS}

In this section, we shall briefly recall the evolution equations
for dipole scattering amplitudes in QCD at high energy and large
$N_c$, together with their physical interpretation in terms of
splitting and merging processes in either the target or the
projectile \cite{IT04,IT05,MSW05}.

The relevant equations describe the rapidity evolution of the
scattering amplitudes $\langle T^{(\kappa)}\rangle$ for the
collision between a system of $\kappa$ dipoles (the `projectile')
and a generic hadronic target. The dipoles making up the
projectile are assumed to be small enough for their scattering to
lie within the scope of perturbation theory. In the course of the
evolution, amplitudes $\langle T^{(\kappa)}\rangle$ corresponding
to different values of $\kappa$ get mixed with each other, via the
gluon splitting and recombination effects. Thus, the equations
form an infinite hierarchy, whose general structure can be easily
appreciated by inspection of the first two equations. These read
\cite{IT05}
    \beq\label{T1evol}
    \frac{\del \lan T(\bm{x},\bm{y})\ran }
    {\del Y}\,=
    \frac{\abar}{2\pi} \int\limits_{\bm{z}}
    \big[\cal{M}_{\bm{x}\bm{y}\bm{z}}
    \otimes \lan T(\bm{x},\bm{y}) \ran
    -\cal{M}(\bm{x},\bm{y},\bm{z})\,
    \lan T^{(2)}(\bm{x},\bm{z};\bm{z},\bm{y}) \ran \big],
    \eeq
and, respectively,
    \beq\label{T2evol}
    \frac{\del
    \lan T^{(2)}(\bm{x}_1,\bm{y}_1;\bm{x}_2,\bm{y}_2)\ran}
    {\del Y}\,=
    &&\frac{\abar}{2\pi} \int\limits_{\bm{z}}
    \Big\{\Big[\cal{M}_{\bm{x}_1\bm{y}_1\bm{z}}
    \otimes \lan T^{(2)}(\bm{x}_1,\bm{y}_1;\bm{x}_2,\bm{y}_2)\ran
    \nonumber \\
    &&-\cal{M}(\bm{x}_1,\bm{y}_1,\bm{z})\,
    \lan T^{(3)}(\bm{x}_1,\bm{z};\bm{z},\bm{y}_1;\bm{x}_2,\bm{y}_2)\ran\Big]
    + \big[1\leftrightarrow 2\,\big]\Big\}
    \nonumber \\
    &&+\left(\frac{\alpha_s}{2\pi}\right)^2
    \frac{\abar}{2 \pi}
    \int\limits_{\bm{u},\bm{v}}
    \cal{R}(\bm{x}_1,\bm{y}_1; \bm{x}_2, \bm{y}_2|\bm{u},\bm{v})\,
    \lan T(\bm{u},\bm{v})\ran,
    \eeq
where $\abar = \alpha_s N_c/\pi$, $Y$ is the rapidity gap between
the projectile and the target, $\lan T(\bm{x},\bm{y})\ran$ is the
scattering amplitude for an external quark--antiquark dipole with
the quark leg at $\bm{x}$ and the antiquark leg at $\bm{y}$,
$\langle T^{(2)}(\bm{x}_1,\bm{y}_1;\bm{x}_2,\bm{y}_2)\rangle$
refers similarly to two external dipoles with coordinates
$(\bm{x}_1,\bm{y}_1)$ and $(\bm{x}_2,\bm{y}_2)$, respectively, and
so forth.

The two kernels $\mathcal{M}$ and $\cal{R}$ which enter the above
equations are defined as follows:
    \beq\label{dipkernel}
    \mathcal{M}(\bm{x},\bm{y},\bm{z})\,\equiv\,
    \frac{(\bm{x}-\bm{y})^2}
    {(\bm{x}-\bm{z})^2 (\bm{y}-\bm{z})^2}\, ,
    \eeq
is the {\em dipole kernel} \cite{AM94,IM031}, which characterizes
the differential probability for an elementary dipole
$(\bm{x},\bm{y})$ to split into two dipoles $(\bm{x},\bm{z})$ and
$(\bm{z},\bm{y})$ per unit rapidity. We have also introduced the
shorthand notation
    \beq\label{Moper}
    \cal{M}_{\bm{x}\bm{y}\bm{z}}
    \otimes f(\bm{x},\bm{y}) \equiv \cal{M}(\bm{x},\bm{y},\bm{z})
    [-f(\bm{x},\bm{y})+f(\bm{x},\bm{z})+f(\bm{z},\bm{y}) ].
    \eeq
Furthermore, $\cal{R}$ is the {\em triple--Pomeron kernel} (this
name will be explained in Sect. \ref{SECT_HS})
    \beq\label{rkernel}
    \cal{R}(\bm{x}_1,\bm{y}_1; \bm{x}_2, \bm{y}_2|\bm{u},\bm{v})
    \,\equiv\,\int\limits_{\bm{z}}
    \lap{u} \lap{v}
    \big[
    \mathcal{M}(\bm{u},\bm{v},\bm{z})\,
    \cal{A}_0(\bm{x}_1,\bm{y}_1|\bm{u},\bm{z})\,
    \cal{A}_0(\bm{x}_2,\bm{y}_2|\bm{z},\bm{v})\,
    \big],
    \eeq
where $\alpha_s^2\,\cal{A}_0$ is (up to a sign) the amplitude for
dipole--dipole scattering in the two--gluon exchange approximation
and for large $N_c$, with
    \beq\label{A0}
    \cal{A}_0(\bm{x},\bm{y}|\bm{u},\bm{v}) \,=\,
    \frac{1}{8}
    \left[\ln \frac{(\bm{x}-\bm{v})^2 (\bm{y}-\bm{u})^2}
    {(\bm{x}-\bm{u})^2 (\bm{y}-\bm{v})^2}
    \right]^2.
    \eeq

The general structure of the higher equations in this hierarchy
can now be easily understood : The amplitude for $\kappa$ dipoles
with $\kappa\ge 2$ is coupled to the one for $\kappa\! -\! 1$
dipoles, to that for $\kappa$ dipoles, and to that for $\kappa \!
+ \!1$ dipoles. For more clarity, in what follows we shall refer
to these three types of terms as the {\em fluctuation terms}, the
(normal) {\em BFKL terms}, and the {\em unitarity corrections},
respectively. (These names should become clear from the subsequent
discussion.) If one ignores the fluctuation terms, which are
formally suppressed by an additional factor of $\alpha_s^2$ (see,
e.g., the last term in Eq.~(\ref{T2evol})), then the above
hierarchy boils down to the large--$N_c$ version of the
Balitsky--JIMWLK hierarchy \cite{B,Janik,LL04}. However, as
pointed out in Refs. \cite{MS04,IMM04,IT04}, the fluctuation terms
are in fact leading--order effects whenever the target is so
dilute (or, equivalently, the external dipole $(\bm{x},\bm{y})$ is
so small) that $\lan T(\bm{x},\bm{y})\ran \simle \alpha_s^2$. Note
also that $T=1$ is a fixed point of the evolution described by
these equations\footnote{This is perhaps more obvious after
rewriting the fluctuation terms as in Eq.~(\ref{T2fluct}).}, which
physically corresponds to the `black disk' limit at high energy.

The above equations are invariant under a Lorentz boost along the
collision axis \cite{BIIT05}, but the physical interpretation of
the various terms depends upon the frame in which we visualize the
evolution. As mentioned in the Introduction, it is always possible
to choose the frame in such a way that a given term corresponds to
{\em dipole splitting}.

Namely, the {\em BFKL terms} and the {\em unitarity corrections}
(that is, all the terms in Eqs.~(\ref{T1evol})--(\ref{T2evol})
except for the last term in Eq.~(\ref{T2evol})) can be interpreted
as dipole splitting in the evolution of the {\em projectile} : If
the rapidity increment $dY$ is given to the projectile, one of the
$\kappa$ original dipoles there splits into two new dipoles, which
then scatter off the target. From this perspective, the BFKL terms
$\sim T^{(\kappa)}$ correspond to the {\em separate} scattering of
the daughter dipoles (plus virtual corrections), whereas the
unitarity effects $\sim T^{(\kappa+1)}$ are rather associated with
their {\em simultaneous} scattering.

Alternatively, the rapidity increment $dY$ can be given to the
{\em target}, in which case it is the {\em fluctuation term} $\sim
T^{(\kappa-1)}$ (e.g., the last term in Eq.~(\ref{T2evol})) which
admits a simple, dipolar, interpretation \cite{IT04}: In the weak
scattering regime where such fluctuations are truly important, the
target is dilute and can be itself described, within the
large--$N_c$ approximation, as a collection of dipoles which
evolve through dipole splitting \cite{AM94,IM031,MSW05}. Then the
last term in Eq.~(\ref{T2evol}) describes the splitting of the
target dipole $(\bm{u},\bm{v})$ into two new dipoles
$(\bm{u},\bm{z})$ and $(\bm{z},\bm{v})$, which then scatter off
the external dipoles $(\bm{x}_1,\bm{y}_1)$ and
$(\bm{x}_2,\bm{y}_2)$ via twice two--gluon exchange. This
interpretation becomes perhaps more transparent after performing
some integrations by parts to rewrite this term as
\begin{align}\label{T2fluct}
    \frac{\partial \lan T^{(2)}(\bm{x}_1,\bm{y}_1;\bm{x}_2,\bm{y}_2)\ran}
    {\partial Y}\,
    \bigg|_{\rm fluct} =
    \left(\frac{\alpha_s}{2\pi}\right)^2
    \frac{\abar}{2 \pi}
    \int\limits_{\bm{u},\bm{v},\bm{z}}&
    \mathcal{M}(\bm{u},\bm{v},\bm{z})\,
    \cal{A}_0(\bm{x}_1,\bm{y}_1|\bm{u},\bm{z})\,
    \cal{A}_0(\bm{x}_2,\bm{y}_2|\bm{z},\bm{v})\,
    \nonumber \\
    &\times\nabla_{\bm{u}}^2 \nabla_{\bm{v}}^2\,
    \lan T(\bm{u},\bm{v})\ran,
 \end{align}
(this is the original form of this term in Ref. \cite{IT05}) and
then observing that, in the dilute regime, $\nabla_{\bm{u}}^2
\nabla_{\bm{v}}^2 \lan T(\bm{u},\bm{v})\ran$ is proportional to
the average dipole number density in the target $\lan
n(\bm{u},\bm{v})\ran$, as it will be explained in Sect.
\ref{SECT_HD} below.

The BFKL terms too can be understood as dipole splitting in the
{\em target} \cite{AM94,IM031}. On the other hand, the unitarity
corrections do not admit a dipolar interpretation in this frame:
From the perspective of target evolution, they rather correspond
to {\em saturation effects}, that is, to {\em gluon recombination}
in the target wavefunction. Similarly, the fluctuation terms can
be associated with gluon recombination in the projectile
wavefunction. We thus arrive at a physical picture in which, in a
given frame, some of the terms in the evolution equations at large
$N_c$ can be directly interpreted as {\em dipole} splitting,
whereas the other terms correspond (in a less direct way, though)
to {\em gluon} merging. As anticipated in the Introduction, and it
will be explained in detail in what follows, the gluon merging is
the process which prevents a fully dipolar interpretation for the
evolution equations even at large $N_c$.

\section{Dipoles in the target}
\setcounter{equation}{0} 
\label{SECT_TARGET}

In this section, we shall demonstrate that
Eqs.~(\ref{T1evol})--(\ref{T2evol}) for the scattering amplitudes
cannot be obtained as the result of a stochastic process involving
dipoles in the target. To that aim, we shall start with the
Hamiltonian formulation of these equations \cite{BIIT05}, which in
the dilute regime reduces to the original dipole picture by
Mueller (in its `color glass' version of Refs.
\cite{IM031,MSW05}), and thus represents a natural starting point
in the search for a dipole interpretation.

\subsection{Hamiltonian formulation in the
$\alpha$--representation} \label{SECT_ALPHA}

\comment{
 there are (at least) two possible
strategies. First, one could try and identify effective `dipoles'
within the structure of the previous equations, and then check
whether the evolution equations obeyed by such `dipoles' are
indeed consistent with the expected equations for a
reaction--diffusion problem. Second, one could follow the reverse
path, namely start by applying a

We now turn to the Hamiltonian formulation of the evolution
equations for the scattering amplitudes, as developed in Refs.
\cite{MSW05,BIIT05}, which has a very suggestive Pomeron structure
and, at a first sight, it seems to support the picture of dipole
splitting and recombination.}

As shown in \cite{BIIT05}, the whole hierarchy of evolution
equations for dipole scattering amplitudes as presented in Sect.
\ref{SECT_TEQS} can be derived from the following effective
Hamiltonian
    \beq\label{Ham}
        H = H_0 + H_{1 \to 2} + H_{2 \to 1},
    \eeq
where the three distinct terms correspond, respectively, to BFKL
evolution, gluon splitting, and gluon merging in the wavefunction
of the target. They are given by\footnote{The splitting piece
$H_{1 \to 2}$ of this Hamiltonian, Eq.~(\ref{MSW}), has been first
obtained by Mueller, Shoshi and Wong \cite{MSW05}. The merging
piece $H_{2 \to 1}$, Eq.~(\ref{H21}), has been introduced in Ref.
\cite{BIIT05}, as the dual counterpart of $H_{1 \to 2}$.} :
    \beq\label{HBFKL}
    H_0=
    \frac{1}{2 N_c^2}\, \atpi
    \int\limits_{\bm{u},\bm{v},\bm{z}}
    \cal{M}(\bm{u},\bm{v},\bm{z})
    \left[ \alpha^a(\bm{u}) -\alpha^a(\bm{z}) \right]
    [\alpha^a(\bm{z}) -\alpha^a(\bm{v})]
    \frac{\delta}{\delta \alpha^b(\bm{u})}
    \frac{\delta}{\delta \alpha^b(\bm{v})},
    \eeq

    \beq\label{MSW}
    H_{1 \to 2} =
    -\frac{g^2}{16 N_c^3}\, \atpi
    \int && \cal{M}(\bm{u},\bm{v},\bm{z})
    \cal{G}(\bm{u}_1|\bm{u},\bm{z})
     \cal{G}(\bm{v}_1|\bm{u},\bm{z})
    \cal{G}(\bm{u}_2|\bm{z},\bm{v})
    \cal{G}(\bm{v}_2|\bm{z},\bm{v})
    \nonumber\\
    && \times
     \nabla_{\bm{u}}^2 \nabla_{\bm{v}}^2
    \alpha^c(\bm{u}) \alpha^c(\bm{v})\,
    \frac{\delta}{\delta \alpha^a(\bm{u}_1)}
    \frac{\delta}{\delta \alpha^a(\bm{v}_1)}
    \frac{\delta}{\delta \alpha^b(\bm{u}_2)}
    \frac{\delta}{\delta \alpha^b(\bm{v}_2)},
    \eeq

    \beq\label{H21}
    H_{2 \to 1}=
    \frac{g^2}{16 N_c^3} \atpi
    \!\int\limits_{\bm{u},\bm{v},\bm{z}}
    \!\!\!\cal{M}(\bm{u},\bm{v},\bm{z})
    \left[ \alpha^a(\bm{u}) -\alpha^a(\bm{z}) \right]^2
    [\alpha^b(\bm{z}) -\alpha^b(\bm{v})]^2
    \frac{\delta}{\delta \alpha^c(\bm{u})}
    \frac{\delta}{\delta \alpha^c(\bm{v})}.\,\,
    \eeq
In these equations, $\alpha^a(\bm{z})$ is the total color gauge
field radiated by color sources within the target (in an
appropriate gauge), and the function
$\cal{G}(\bm{u}_1|\bm{u},\bm{z})$ represents (up to a factor
$g\,t^a$) the classical field created at $\bm{u}_1$ by the
elementary dipole $(\bm{u},\bm{z})$:
    \beq\label{calG}
    \cal{G}(\bm{u}_1|\bm{u},\bm{z}) =
    \frac{1}{4\pi}
    \ln \frac{(\bm{u}_1-\bm{z})^2}{(\bm{u}_1-\bm{u})^2}.
    \eeq
In what follows, we shall refer to the above form of the effective
theory as the ``$\alpha$--representation''. The operators
$\alpha^a(\bm{z})$ and $\delta/\delta\alpha^a(\bm{z})$ can be
interpreted as `Fock--space operators' which annihilate and create
a gluon, respectively. This makes it clear that the above
Hamiltonian describes the evolution of the gluon distribution in
the target, which proceeds through BFKL evolution ($H_0$), $2 \to
4$ gluon splitting ($H_{1 \to 2}$), and $4 \to 2$ gluon merging
($H_{2 \to 1}$). From this perspective, the splitting  Hamiltonian
(\ref{MSW}) \cite{MSW05} should be more naturally denoted as $H_{2
\to 4}$ (and similarly $H_{4 \to 2}$ for the merging term
(\ref{H21})), but the present notations anticipate the `dipolar'
form of the Hamiltonian, to be presented later. The Hamiltonian in
Eqs.~(\ref{Ham})--(\ref{H21}) has a remarkable {\em self--duality}
property \cite{KL3,BIIT05}, to be discussed in Sect.
\ref{SECT_HS}.


In this representation, the scattering amplitudes for external
dipoles are obtained as $\langle T^{(\kappa)}\rangle = \langle\,
T(1)T(2) ... T(\kappa)\rangle$, where $T\equiv T(\bm{x},\bm{y})$
is the following scattering operator:
    \beq\label{Tdipole0}
    T(\bm{x},\bm{y}) =
    \frac{g^2}{4 N_c}\,
    \left[\alpha^a(\bm{x})-\alpha^a(\bm{y})\right]^2,
    \eeq
which appears to describe single scattering in the two--gluon
exchange approximation. However, within the present formalism,
Eq.~(\ref{Tdipole0}) is to be used in the whole kinematical range,
including at high energy where the multiple scattering is
important. This is appropriate since the $\alpha$--representation
described here is only an {\em effective} theory, in which the
various microscopic process are not always faithfully depicted,
but which reproduces, by construction, the correct evolution
equations for the scattering amplitudes (that is, the hierarchy
starting with Eqs.~(\ref{T1evol})--(\ref{T2evol})). Indeed, by
using the above formul\ae, it is straightforward to check that the
Hamiltonian evolution equations
    \beq\label{HaonT}
        \frac{\del \lan T^{(\kappa)} \ran}{\del Y} =
        \lan H\, T^{(\kappa)} \ran
    \eeq
generate the expected hierarchy, so long as one neglects
contractions which give terms suppressed at large--$N_c$. For
example, when acting with $\delta^2/\delta\alpha^a_{\bm{u}_1}
\delta\alpha^a_{\bm{v}_1}$ from $H_{1 \to 2}$ on the two--dipole
scattering operator
$T^{(2)}(\bm{x}_1,\bm{y}_1;\bm{x}_2,\bm{y}_2)$, one should keep in
the equation only the terms which are generated when both
derivatives act on the same dipole. In that sense, the action of
the Hamiltonian is constrained by a large--$N_c$ counting rule.
Moreover, the  Hilbert space for the $\alpha$--representation is
also constrained: Because of the assumptions underlying the
construction of the Hamiltonian (\ref{Ham})--(\ref{H21}) (see the
discussion in Ref. \cite{BIIT05}), this Hamiltonian cannot be used
to act on {\em arbitrary} operators built with $\alpha$, but only
on the operators built with the scattering amplitude
(\ref{Tdipole0}). (Such acceptable operators are gauge invariant
in the sense explained in Ref. \cite{ODDERON}.)

\subsection{Hamiltonian theory in the $D$--representation}
\label{SECT_HD}

In the $\alpha$--representation introduced above, the dipolar
structure of the Hamiltonian is not explicit. But as we shall
explain below, this structure becomes manifest after a change of
representation, which has also the advantage to eliminate all the
constraints on the Hamiltonian structure that we have previously
mentioned.

Specifically, one can reexpress the Hamiltonian and the
interesting observables (the scattering amplitudes) in terms of
operators which are {\em bilinear} in the number of fields
$\alpha$ or of functional derivatives $\delta/\delta \alpha$. If
$\rho_a$ denotes the color charge density in the target, so that
    \beq\label{Poisson}
    -\lap{x} \alpha_a(\bm{x}) =  \rho_a(\bm{x}),
    \eeq
then we define the $D$--representation via the following
identifications:
    \beq\label{defd}
    D_{\bm{x}\bm{y}} = -\frac{1}{g^2 N_c}\, \rho_a(\bm{x}) \rho_a(\bm{y})
    \quad, \quad \bm{x} \neq \bm{y}
    \eeq

    \beq\label{defdd}
    D_{\bm{x}\bm{y}}^{\dagger} = 1 + \frac{g^2}{4 N_c}
    \left[
    \frac{\delta}{\delta \rho^a(\bm{x})}-\frac{\delta}{\delta \rho^a(\bm{y})}
    \right]^2.
    \eeq
The notation is slightly abusive, since $D$ and $D^{\dagger}$ are
not Hermitian conjugate to each other in any sense, but it is
suggestive of the fact that these operators act as creation and
annihilation operators for dipoles in the target (see Sect.
\ref{SECT_TDIP} below).

When expressed in terms of $D$ and $D^{\dagger}$, the Hamiltonian
takes a rather compact form:
    \beq\label{HamD}
    H\left[ D,D^{\dagger}\right] =&&
    \frac{\abar}{2 \pi}
    \int\limits_{\bm{u},\bm{v},\bm{z}}
    \cal{M}(\bm{u},\bm{v},\bm{z})\,D_{\bm{u}\bm{v}}
    \left[
    -D^{\dagger}_{\bm{u}\bm{v}}
    + D^{\dagger}_{\bm{u}\bm{z}} D^{\dagger}_{\bm{z}\bm{v}}
    \right]
    \nn
    &&- \frac{1}{2}\left(\frac{\alpha_s}{2\pi}\right)^2 \atpi
    \int
    \cal{R}(\bm{u}_1,\bm{v}_1; \bm{u}_2, \bm{v}_2|\bm{u},\bm{v})\,
    D_{\bm{u}_1\bm{v}_1}
    D_{\bm{u}_2\bm{v}_2}D^{\dagger}_{\bm{u}\bm{v}}\,,
    \eeq
where the expression in the first line has been obtained by
summing up the BFKL\footnote{There is in fact a slight subtlety
concerning the BFKL Hamiltonian, which in going from
Eq.~(\ref{HBFKL}) to Eq.~(\ref{HamD}) has been rewritten in a
different, but equivalent, form; see the discussion in Ref.
\cite{BIIT05}.} and the splitting pieces of the Hamiltonian, cf.
Eqs.~(\ref{HBFKL})--(\ref{MSW}), while that in the second line is
a rewriting of the merging Hamiltonian, Eq.~(\ref{H21}).

Furthermore, the dipole scattering amplitude can be related to the
operator $D_{\bm{x}\bm{y}}$ via
    \beq\label{TD}
    T(\bm{x},\bm{y}) [D]\,=\,\alpha_s^2
    \int \limits_{\bm{u},\bm{v}}
    \cal{A}_0(\bm{x},\bm{y}|\bm{u},\bm{v}) D_{\bm{u}\bm{v}}\,,
    \eeq
which follows from Eqs.~(\ref{Tdipole0}), (\ref{Poisson}), and
(\ref{A0}), together with the condition that the system be
globally color neutral: $\int_{\bm{u}} \rho^a({\bm{u}}) = 0$.

To complete the definition of the new representation, one still
needs to specify the algebra of the operators $D$ and
$D^{\dagger}$. To that aim, we rely on their respective
expressions in terms of $\rho$ and $\delta/\delta \rho$, cf.
Eqs.~(\ref{defd})--(\ref{defdd}), to conclude that, in the
representation in which $D_{\bm{x}\bm{y}}$ is diagonal (i.e., a
pure function) and in the large--$N_c$ approximation,
$D^{\dagger}$ can be represented as a functional derivative with
respect to $D$ :
 \beq\label{DDDeltdD}
 D_{\bm{x}\bm{y}}^{\dagger} = 1 + \frac{\delta}{\delta
 D_{\bm{x}\bm{y}}}\,,\qquad \mbox{with}\qquad \frac{\delta}{\delta
 D_{\bm{x}\bm{y}}}\,D_{\bm{u}\bm{v}}\,\equiv\,
 \frac{1}{2} \left(
    \delta_{\bm{x}\bm{u}} \delta_{\bm{y}\bm{v}}+
    \delta_{\bm{x}\bm{v}}\delta_{\bm{y}\bm{u}}
    \right).
    \eeq
To understand how this arises, note first that, for $\bm{u} \neq
\bm{v}$,
 \beq\label{COMMUT}
 \frac{g^2}{4 N_c}
    \left[
    \frac{\delta}{\delta \rho^a(\bm{x})}-\frac{\delta}{\delta \rho^a(\bm{y})}
    \right]^2 \, \left( -\frac{1}{g^2 N_c}\,
    \rho_a(\bm{u}) \rho_a(\bm{v})\right)
 \,\simeq\,\frac{1}{2} \left(
    \delta_{\bm{x}\bm{u}} \delta_{\bm{y}\bm{v}}+
    \delta_{\bm{x}\bm{v}}\delta_{\bm{y}\bm{u}}
    \right),
    \eeq
where we have used $N_c^2-1\simeq N_c^2$ at large $N_c$.
Furthermore, if we temporarily return to the
$\rho$--representation in Eqs.~(\ref{defd})--(\ref{defdd}) and
consider the action of $D^{\dagger}$ on a string of
$D$--operators, so like $D_{\bm{x}\bm{y}}^{\dagger}
D_{\bm{u}_1\bm{v}_1}D_{\bm{u}_2\bm{v}_2}\dots$, then it is clear
that  the dominant result at large $N_c$ is the same as obtained
when acting with the functional derivative ${\delta}/{\delta
D_{\bm{x}\bm{y}}}$.

Thus, by using the expression (\ref{HamD}) for the Hamiltonian
together with Eq.~(\ref{DDDeltdD}) for $D^{\dagger}$, one ensures
that the action of $H$ on operators built with $D$ generates only
the would--be dominant contributions at large--$N_c$. Moreover,
the Hilbert space in the $D$--representation is not constrained
anymore: Indeed, Eq.~(\ref{TD}) can be inverted to yield
\beq\label{invert} D_{\bm{x}\bm{y}}
    = \frac{2}{g^4}
    \nabla_{\bm{x}}^2 \nabla_{\bm{y}}^2\, T(\bm{x},\bm{y})
    \qquad {\rm for} \qquad
    \bm{x}\neq \bm{y},
 \eeq
showing that any operator built with $D$ is physically acceptable.
Thus, as anticipated, by changing from the $\alpha$ to the
$D$--representation we have eliminated all the constraints.

It is now a straightforward exercise to verify that the evolution
equations for $\langle D^{(\kappa)} \rangle \equiv \langle
D(1)D(2)\dots D(\kappa)\rangle$ generated as:
    \beq\label{HDonD}
        \frac{\del \langle D^{(\kappa)} \rangle}{\del Y} =
        \lan H\, D^{(\kappa)} \ran
    \eeq
lead to the expected equations for the scattering amplitudes after
also using Eq.~(\ref{TD}). 

\comment{One should perhaps notice at this point that the merging
piece in the Hamiltonian (\ref{HamD}), which involves the kernel
${\cal R}$, generates the unitarity corrections in the equations
for the scattering amplitudes, and not the fluctuation terms there
! The latter, together with the normal BFKL terms, are rather
induced by the splitting piece in the first line of
Eq.~(\ref{HamD}).}

\subsection{A dipolar picture and its limitations}
\label{SECT_TDIP}

We now turn to the dipolar interpretation of the evolution
described by the effective Hamiltonian (\ref{HamD}). As we shall
soon discover, such an interpretation is truly meaningful (in the
sense of having a probabilistic meaning) only in the {\em dilute}
regime, where it reduces to the standard dipole picture by Mueller
\cite{AM94,AM95}. Namely, for a dilute target, the recombination
effects can be ignored and the Hamiltonian (\ref{HamD}) reduces to
the expression in its first line, i.e., $H_{\rm dipole}\equiv H_0
+ H_{1 \to 2}$, which is recognized as the `color glass' version
of the Hamiltonian generating the dipole picture
\cite{IM031,MSW05}. As we briefly recall now, the evolution driven
by $H_{\rm dipole}$ leads to a probabilistic picture of the
target, as a collection of dipoles which evolves through splitting
(see Refs. \cite{AM94,AM95,IM031,MSW05,HIMS05} for more details).

Specifically, in the dilute regime, the operators $D^{\dagger}$
and $D$ introduced before can be interpreted as dipole creation
and annihilation operators, respectively, with the following
commutation relation (at large $N_c$):
 \beq\label{Dcom}
    \left[
    D_{\bm{x}\bm{y}},D^{\dagger}_{\bm{u}\bm{v}}
    \right]=
    \frac{1}{2} \left(
    \delta_{\bm{x}\bm{u}} \delta_{\bm{y}\bm{v}}+
    \delta_{\bm{x}\bm{v}}\delta_{\bm{y}\bm{u}}
    \right).
    \eeq
The target distribution is then obtained by successively acting
with the creation operators on the `dipole vacuum', and can be
represented as the following `color glass weight function' :
 \beq \label{ZDM}
 Z_Y[\rho ] \; = \;
 \sum_{N=1}^{\infty} \, \int d\Gamma_N \,
 P_N(\{\bm{u}_i,\bm{v}_i\};Y)
 \, \prod_{i=1}^{N} D^{\dagger}(\bm{u}_{i},\bm{v}_i)
  \, \delta[\rho] \; , \eeq
where the functional distribution $\delta[\rho]$ represents the
vacuum (no color charge, and hence no dipole !),
$Z_N(\{\bm{u}_i,\bm{v}_i\}) \equiv \prod_{i=1}^{N}
D^{\dagger}(\bm{u}_{i},\bm{v}_i) \delta[\rho]$ describes a given
configuration of $N$ dipoles, $P_N(\{\bm{u}_i,\bm{v}_i\};Y)$ is
the corresponding probability at rapidity $Y$, and $d\Gamma_N$
denotes the measure for the $N$--dipole phase--space: $d{\Gamma}_N
= (1/N!)\prod_{i=1}^{N}{\rm d}^2\bm{u}_i{\rm d}^2\bm{v}_i$.

The `weight function' (\ref{ZDM}) acts as a probability
distribution for $\rho_a$ when computing target expectation
values. For instance:
  \beq\label{AVD}
 \lan D(\bm{x},\bm{y})\ran &= & \int D[\rho]\,
 D(\bm{x},\bm{y})\,Z_{Y}[\rho]\,\nn &= &\,
 \sum_{N=1}^{\infty} \, \int d\Gamma_N \, P_N(\{\bm{u}_i,\bm{v}_i\};Y)
    \sum_{i=1}^N
 \delta_{\bm{u}_{i}\bm{x}}\delta_{\bm{v}_{i}\bm{y}}\,
 \equiv \, \lan n(\bm{x},\bm{y})\ran,\eeq
where in writing the second line we have used Eqs.~(\ref{Dcom})
and (\ref{ZDM}), and then recognized the definition of the {\em
average dipole number density}.

In view of the discussion above, the dipole interpretation of the
evolution generated by $H_{\rm dipole}$ is quite transparent (cf.
the first line of Eq.~(\ref{HamD})) : The term involving
$D_{\bm{u}\bm{v}} D^{\dagger}_{\bm{u}\bm{z}}
D^{\dagger}_{\bm{z}\bm{v}}$ describes the splitting of the
original dipole $({\bm{u},\bm{v}})$ into two new dipoles
$({\bm{u},\bm{z}})$ and $({\bm{z},\bm{v}})$, with a probability
density $({\abar}/{2 \pi})\cal{M}(\bm{u},\bm{v},\bm{z})$ per unit
of phase--space and per unit rapidity. The other
term, proportional to $
-D_{\bm{u}\bm{v}} D^{\dagger}_{\bm{u}\bm{v}}$, describes the
decrease in the probability that the original dipole remain
unchanged after one step in the evolution.

To verify that this probabilistic interpretation is mathematically
well founded, one can use $H_{\rm dipole}$ to deduce evolution
equations for either the probability densities $P_N$, or the
dipole number densities $\langle n^{(\kappa)} \rangle \equiv
\langle D^{(\kappa)} \rangle$, and then check that the ensuing
equations have the structure expected for the Master equations
associated with a stochastic process. For the splitting problem
described by $H_{\rm dipole}$, both tests have been performed in
the literature \cite{Salam95,IM031,IT04}, and they confirm the
probabilistic interpretation given above. In what follows we shall
follow the second strategy, namely, the one which focuses on the
particle densities, but we shall do so in the more general context
of an evolution involving {\em both splitting and merging}, as
generated by the full Hamiltonian in Eq.~(\ref{HamD}).

Indeed, a question which naturally arises at this point is whether
one can extend this dipole picture to the high--density regime as
well, that is, to the merging piece of the Hamiltonian, as given
in the second line of Eq.~(\ref{HamD}). Since proportional to
$D_{\bm{u}_1\bm{v}_1}
D_{\bm{u}_2\bm{v}_2}D^{\dagger}_{\bm{u}\bm{v}}$, is seems natural
to interpret this piece as describing the recombination of two
dipoles $({\bm{u}_1,\bm{v}_1})$ and $({\bm{u}_2,\bm{v}_2})$ into a
single dipole $({\bm{u},\bm{v}})$, with a probability density
which is proportional to $\cal{R}(\bm{u}_1,\bm{v}_1; \bm{u}_2,
\bm{v}_2|\bm{u},\bm{v})$ (and with an overall sign which is not
clear yet; this will be clarified in Sect. \ref{SEC_MMERG}).
However, a more careful analysis reveals that this interpretation
is in fact deceiving, as shown by the following argument: For the
probabilistic interpretation to be meaningful, the
`triple--Pomeron' kernel $\cal{R}$ must have a definite sign
(since a probability density cannot change sign !) However, a
brief inspection of Eq.~(\ref{rkernel}) reveals that the actual
kernel $\cal{R}$ has {\em not} a fixed sign. Indeed, using the
fact that this kernel is a total derivative and that there are no
boundary terms when it is integrated over all the transverse space
(as we shall shortly check), we immediately find
    \beq\label{rint}
    \int\limits_{\bm{u},\bm{v}}
    \cal{R}(\bm{u}_1,\bm{v}_1; \bm{u}_2, \bm{v}_2|\bm{u},\bm{v})
    =0,
    \eeq
showing that the value of the vertex must be negative in some
regions of the transverse space, and positive in some others.
Clearly, this property prohibits any probabilistic interpretation
for the merging vertex in the Hamiltonian (\ref{HamD}).

At this point, we should stress that the aforementioned sign
property of the triple--Pomeron kernel does not entail any sign
problem for the fluctuation term in Eq.~(\ref{T2evol}), and thus
for the actual evolution equations in QCD: Even though $\cal{R}$
has no definite sign, the convolution yielding the last term in
Eq.~(\ref{T2evol}) is nevertheless positive--definite within the
whole kinematical range in which this term is important, that is,
in the weak scattering regime. To verify this, it is preferable to
rewrite this term as in Eq.~(\ref{T2fluct}) and then use the fact,
for a dilute target, the quantity  $\nabla_{\bm{u}}^2
\nabla_{\bm{v}}^2 \lan T(\bm{u},\bm{v})\ran$ is
positive--semidefinite, since proportional to the average dipole
number density (cf. Eqs.~(\ref{invert}) and (\ref{AVD})).

To render the previous arguments more precise, we shall compare in
Sect. \ref{SEC_MASTER} the equations for the dipole number
densities generated by the Hamiltonian (\ref{HamD}) to those
arising from an actual reaction--diffusion process with generic
vertices for splitting and recombination. To that aim, let us
derive here the corresponding equations in QCD. The `dipole number
density' is defined as the expectation value of $D$, cf.
Eq.~(\ref{AVD}), and similarly for the higher $\kappa$--body
densities (e.g., $\lan D D\ran$ is the average {\em pair} density,
etc.). One should keep in mind, though, that such expectation
values can be meaningfully interpreted as particle number
densities only in the dilute regime. The corresponding equations
are then obtained according to Eq.~(\ref{HDonD}). It is convenient
to separate the respective contributions of the splitting ($H_0 +
H_{1 \to 2}$) and merging ($H_{2 \to 1}$) terms in the
Hamiltonian.

For the BFKL+splitting contributions, one finds:
   \beq\label{D1evol}
    \frac{\del \lan D_{\bm{u}\bm{v}}\ran}
    {\del Y}\Big|_{1\to 2}\,=
    \frac{\abar}{2\pi} \int\limits_{\bm{z}}
    \cal{K}_{\bm{u}\bm{v}\bm{z}}
    \otimes \lan D_{\bm{u}\bm{v}} \ran,
    \eeq
and furthermore
    \beq\label{D2evol}
    \frac{\del \lan D_{\bm{u}_1\bm{v}_1} D_{\bm{u}_2\bm{v}_2}\ran}
    {\del Y}\Big|_{1\to 2}\,=
    \frac{\abar}{2\pi} \Big[\int\limits_{\bm{z}}&&
    \cal{K}_{\bm{u}_1\bm{v}_1\bm{z}}
    \otimes \lan  D_{\bm{u}_1\bm{v}_1} D_{\bm{u}_2\bm{v}_2}\ran
    \nn
    &&+\delta(\bm{u}_2-\bm{v}_1)
    \cal{M}(\bm{u}_1,\bm{v}_2,\bm{u}_2)
    \lan D_{\bm{u}_1\bm{v}_2} \ran \Big] +
    \big\{1 \leftrightarrow 2 \big\} ,
    \eeq
where he have used the shorthand notation
    \beq\label{Koper}
    \cal{K}_{\bm{u}\bm{v}\bm{z}}
    \otimes f(\bm{u},\bm{v}) \equiv
    &&-\cal{M}(\bm{u},\bm{v},\bm{z}) f(\bm{u},\bm{v})
    +\cal{M}(\bm{u},\bm{z},\bm{v}) f(\bm{u},\bm{z})
    \nn
    &&+\cal{M}(\bm{z},\bm{v},\bm{u}) f(\bm{z},\bm{v}).
    \eeq
Note, in particular, the term linear in $\lan D\ran$ in the r.h.s.
of Eq.~(\ref{D2evol}) for $\langle D^{(2)}\rangle$. This term
describes a `fluctuation' in which two dipoles which have a common
leg are produced in one step of the evolution, via the splitting
of a single original dipole. Through Eq.~(\ref{TD}), this
mechanism produces the fluctuation term in Eq.~(\ref{T2evol}).

The recombination piece in the equation for $\lan D\ran$ is
similarly obtained as
    \beq\label{D1evolR}
    \frac{\del \lan D_{\bm{u}\bm{v}}\ran}
    {\del Y}\Big|_{2 \to 1}\,=
    -\left(\frac{\alpha_s}{2\pi}\right)^2
    \frac{\abar}{2\pi} \int\limits_{\bm{u}_i,\bm{v}_i}
    \cal{R}(\bm{u}_1,\bm{v}_1; \bm{u}_2, \bm{v}_2|\bm{u},\bm{v})
    \lan D_{\bm{u}_1\bm{v}_1} D_{\bm{u}_2\bm{v}_2}\ran.
    \eeq
Via Eq.~(\ref{TD}), this piece produces the non--linear term $\sim
\langle T^{(2)}\rangle$ in the r.h.s. of Eq.~(\ref{T1evol}). The
higher equations in the hierarchy for $\langle
D^{(\kappa)}\rangle$ can be derived similarly, but the equations
written above will in fact suffice to illustrate our point in the
subsequent discussion.

To conclude this section, let us explicitly check the fact that
the surface terms for the integral (\ref{rint}) vanish indeed, as
announced. Concentrating, for example, in the $\bm{u}$ integration
there, we have
    \beq\label{rint1}
    \int\limits_{\bm{u}}
    \cal{R}(\bm{u}_1,\bm{v}_1; \bm{u}_2, \bm{v}_2|\bm{u},\bm{v})
    \equiv \int\limits_{\bm{u}}
    \lap{u} F(\bm{u})
    = \lim_{R \to \infty}
    \int\limits_{\mathcal{C}_R}
    \dif \bm{\Sigma}
    \!\cdot\!
    \grad_{\bm{u}} F(\bm{u}),
    \eeq
with the one--dimensional ``surface'' integral to be taken along
$\mathcal{C}_R$, the circle of radius $R$. More precisely
    \beq\label{dsigma}
    \dif \bm{\Sigma} = u \,\dif \phi \,\hat{\bm{u}},
    \eeq
where $\phi$ is the polar angle of the vector $\bm{u}$,
$\hat{\bm{u}}$ is the unit vector along the radial direction, and
$u \equiv |\bm{u}|$ is equal to $R$ on $\mathcal{C}_R$. By
expanding the dipole kernel $\mathcal{M}(\bm{u},\bm{v},\bm{z})$
and the dipole--dipole scattering
$\cal{A}_0(\bm{u}_1,\bm{v}_1|\bm{u},\bm{z})$ for large values of
$u$ it is straightforward to show that
    \beq\label{Fexpand}
    F(\bm{u}) = A + \frac{\bm{B}\!\cdot\! \bm{u}}{u^2} +
    \dots
    \Longrightarrow
    \grad_{\bm{u}} F(\bm{u}) =
    -\frac{B}{u^2}
    \left(\cos \phi \,\hat{\bm{u}}
    + \sin \phi \,\hat{\bm{\phi}}\right)+
    \dots,
    \eeq
with the scalar $A$ and the vector $\bm{B}$ being independent of
$\bm{u}$. Since we are going to integrate over $\phi$, we have
assumed, without any loss of generality, that the polar angle of
the vector $\bm{B}$ is zero. Now it is trivial to see that the
insertion of Eqs.~(\ref{dsigma}) and (\ref{Fexpand}) into
Eq.~(\ref{rint1}) gives a result which vanishes indeed when $R \to
\infty$ :
    \beq\label{rint2}
    \int\limits_{\bm{u}}
    \cal{R}(\bm{u}_1,\bm{v}_1; \bm{u}_2, \bm{v}_2|\bm{u},\bm{v})
    = - B \int\limits_{0}^{2 \pi} \dif \phi \cos \phi\,
    \lim_{R \to \infty} \frac{1}{R} = 0.
    \eeq

\section{The general reaction--diffusion process}
\setcounter{equation}{0} \label{SEC_MASTER}

In this section, we shall describe a generic reaction--diffusion
process with non--local vertices for particle (`dipole') splitting
and recombination. Starting from the Master equation for
probabilities, we shall deduce an hierarchy of evolution equations
for the particle number density and its correlations. For more
clarity, we shall discuss separately the effects of splitting and
merging, which is indeed possible since the two types of processes
take place independently from each other. (In the complete
equations, which are displayed for convenience in the Appendix,
the corresponding terms simply add up to each other.) This is in
line with our general strategy which is to emphasize the disparity
(in so far as the dipolar interpretation is concerned) between
splitting and merging contributions to the actual evolution
equations in QCD.

\subsection{Master equation for splitting}
\label{SEC_MSPLIT}

To facilitate the correspondence with the QCD problem at hand, we
shall use a QCD--inspired terminology also for the generic
reaction--diffusion problem. The system is therefore a `hadron'
whose general state is characterized by the number of `dipoles'
$N$ together with the associated coordinates $(u_1 \,...\, u_N )$,
where $u_i$ is a 4--dimensional variable representing in a
collective way the coordinates of the quark and the antiquark of
the $i$--th dipole. Now, let us assume that, under a step $\dif Y$
in `time' (or `rapidity'), a dipole can split into two
non--locally. The corresponding Master equation is
    \beq\label{mastera}
        \frac{\del P_N}{\del Y} \Big|_{\alpha} =
        \frac{\del P_N^{(g)}}{\del Y} \Big|_{\alpha}
        +\frac{\del P_N^{(\ell)}}{\del Y} \Big|_{\alpha},
    \eeq
where $P_N\equiv P_N(u_1 ... u_N)$ is the probability density for
a given configuration of $N$ dipoles (this is totally symmetric
under the permutations of the dipoles) and the superscripts ($g$)
and ($\ell$) stand for the gain and loss terms, respectively. In
their most generic form, the two terms on the r.h.s. of
Eq.~(\ref{mastera}) are given by
    \beq\label{pga}
    \frac{\del P_N^{(g)}(u_1 ... u_N)}{\del Y} \Big|_{\alpha}=
    \sum\limits_{i \neq j} \int\limits_{v}
    \alpha(v|u_i u_j) \,P_{N-1}(u_1...\slashed{u}_i...\slashed{u}_j... u_N
    v),
    \eeq
and, respectively,
    \beq\label{pla}
    \frac{\del P_N^{(l)}(u_1 ... u_N)}{\del Y} \Big|_{\alpha}=-
    \sum\limits_{i} \int\limits_{vw}
    \alpha(u_i|vw) \,P_N(u_1...u_N),
    \eeq
with $\alpha(v|u_i u_j)=\alpha(v|u_j u_i)$ the differential
probability that the dipole $v$ split into the dipoles $u_i$ and
$u_j$ (this is necessarily positive semi--definite, and symmetric
in the coordinates of the child dipoles), and where it is
understood that a slashed variable is to be excluded. It is
straightforward to show that Eqs.~(\ref{mastera})--(\ref{pla})
conserve the total probability, i.e.
    \beq\label{probcons}
    \frac{\del}{\del Y}
    \sum\limits_{N=1}^{\infty}
    \int\dif \Gamma_N
    P_N(u_1...u_N) =0,
    \eeq
where the differential space of integration is
    \beq\label{dgamma}
    \dif \Gamma_N =
    \frac{1}{N!}\,\dif u_1...\dif u_N
    \equiv \frac{\dif \bar{\Gamma}_N}{N!}.
    \eeq
(The factor $1/N!$ is needed to divide out the number of
equivalent configurations). In fact the probability conservation
property is more stringent, as it holds already before summing
over $N$ :
    \beq\label{sprobcons}
    \int \dif \Gamma_N\,
    \frac{\del P_N^{(\ell)}}{\del Y} \Big|_{\alpha} +
    \int \dif \Gamma_{N+1}\,
    \frac{\del P_{N+1}^{(g)}}{\del Y} \Big|_{\alpha} =0.
    \eeq

Using the above Master equation, we would like to derive the
evolution equation for a generic operator $\cal{O}$ which depends
upon the transverse coordinates of the dipoles. The expectation
value of such an operator may be written as
    \beq\label{Oave}
    \lan \cal{O} \ran =
    \sum_{N=1}^{\infty} \int \dif \Gamma_N
    P_N(u_1...u_N)\,
    \cal{O}_N(u_1...u_N).
    \eeq
By using the Master equation together with the assumed permutation
symmetries for $P_N$ and for the splitting vertex, we find
    \beq\label{Oevol}
    \frac{\del \lan \cal{O} \ran}{\del Y} \Big|_{\alpha} =
    \sum\limits_{N=1}^{\infty}
    \frac{\dif \bar{\Gamma}_{N+2}}{(N\!-\!1)!}\,
    &&\alpha(u_N|u_{N+1}u_{N+2})
    P_N(u_1...u_N)
    \nn
    &&\times \big[
    -\cal{O}_N(u_1...u_N)
    +\cal{O}_{N+1}(u_1...u_{N-1}u_{N+1}u_{N+2})
    \big].
    \eeq
To be more specific let us first write the equation which describes
the evolution of the average dipole density. According to the
general definition (\ref{Oave}), this is defined as
    \beq\label{nave}
    \lan n(v) \ran =
    \sum_{N=1}^{\infty} \int \dif \Gamma_N
    P_N(u_1...u_N)\,
    n_N(u_1...u_N),
    \eeq
where $n_N$ is specified to be
    \beq\label{nn}
    n_N = \sum\limits_{i=1}^{N}
    \delta(u_i-v).
    \eeq
Then it is easy to see that only three of the $\delta$-functions
survive in the square bracket in Eq.~(\ref{Oevol}). These give
rise to three terms in the equation for $\lan n(v) \ran$ and we
arrive at
    \beq\label{nevol}
    \frac{\del \lan n(v) \ran}{\del Y}\Big|_{\alpha}=
    \int\limits_{w_1w_2}
    \big[
    2\alpha(w_1|vw_2)\lan n(w_1) \ran
    -\alpha(v|w_1w_2)\lan n(v) \ran
    \big],
    \eeq
where the factor of 2 in the first term comes from the symmetry
property of the vertex $\alpha(v|w_1w_2)=\alpha(v|w_2w_1)$. The
interpretation of the r.h.s. of Eq.~(\ref{nevol}) is transparent:
there is a positive contribution, since the arbitrary dipole $w_1$
can split into the dipoles $v$ and $w_2$, but the latter is not
``measured'' though, and there is a negative contribution, since
the dipole $v$ can be lost by splitting into $w_1$ and $w_2$.

Similarly, the dipole--pair density is defined through
    \beq\label{nn2}
    n_N^{(2)} = \sum\limits_{i \neq j}
    \delta(u_i-v_1) \delta(u_j-v_2),
    \eeq
and evolves according to
    \beq\label{n2evol}
    \frac{\del \lan n^{(2)}(v_1v_2) \ran}{\del Y}\Big|_{\alpha}=
    && \int\limits_{w_1w_2}
    \left[
    2\alpha(w_1|v_1w_2)\lan n^{(2)}(w_1 v_2) \ran
    -\alpha(v_1|w_1w_2)\lan n^{(2)}(v_1v_2) \ran
    \right]
    \nn
    &&+\, \big\{v_1 \leftrightarrow v_2 \big\}
    \,+ \,2\int\limits_w \alpha(w|v_1v_2) \lan n(w) \ran
    .
    \eeq
All the terms in the r.h.s. except for the last one are the
straightforward generalization of the corresponding terms in
Eq.~(\ref{nevol}). As for the last term in Eq.~(\ref{n2evol}),
this describes a `fluctuation' in which an arbitrary dipole $w$
splits into two dipoles $v_1$ and $v_2$ which are both measured.

It is now easy to understand the general structure of the
hierarchy corresponding to splitting: The r.h.s. of the equation
for $\langle n^{(\kappa)} \rangle$ (the $\kappa$--dipole density,
defined by analogy to Eq.~(\ref{nn2})) involves terms proportional
to $\langle n^{(\kappa)} \rangle$, associated to splitting
processes in which only one of the child dipoles is measured, and
terms proportional to $\langle n^{(\kappa-1)} \rangle$, in which
both child dipoles are measured. The latter are the `fluctuation
terms'.

It is furthermore straightforward to check that the above
equations for $\lan n \ran$ and $\langle n^{(2)} \rangle$ will
reduce to Eqs.~(\ref{D1evol}) and (\ref{D2evol}) of the QCD
problem for the specific vertex
  \beq\label{atoM}
    \alpha(v|w_1w_2) &&=
    \alpha_{3{\mathbb P}}(\bm{v}\bar{\bm{v}}|
    \bm{w}_1 \bar{\bm{w}}_1;\bm{w}_2 \bar{\bm{w}}_2)
    \nn
    &&\equiv\frac{1}{2}\,\atpi\,
    \delta(\bm{v}-\bm{w}_1)
    \delta(\bar{\bm{v}}-\bar{\bm{w}}_2)
    \delta(\bar{\bm{w}}_1-\bm{w}_2)
    \cal{M}(\bm{v},\bar{\bm{v}},\bar{\bm{w}}_1)\, + \,
    \big\{1 \leftrightarrow 2 \big\}\,,
    \eeq
(the subscript ``$3{\mathbb P}$'' will be explained in Sect.
\ref{SECT_HS}) where we have separated each 4--dimensional
variable into two 2--dimensional ones, denoting the coordinates of
the quark and the antiquark, respectively. Therefore, the
splitting-in-the-target processes in QCD, as described by the
first line in the effective Hamiltonian (\ref{HamD}), have an
equivalent description in terms of a Master equation, and thus a
well--defined (dipolar) probabilistic interpretation, as
anticipated. The same holds therefore for the BFKL+fluctuation
terms in the equations for the scattering amplitudes, cf.
Eqs.~(\ref{T1evol})--(\ref{T2evol}), when these are viewed from
the perspective of target evolution.

\subsection{Master equation for recombination}
\label{SEC_MMERG}

Let us now similarly consider the master equation for a `dipole'
system which evolves only through recombination. In analogy with
Eqs.~(\ref{mastera})--(\ref{pla}) we can write
    \beq\label{masterab}
        \frac{\del P_N}{\del Y} \Big|_{\beta} =
        \frac{\del P_N^{(g)}}{\del Y} \Big|_{\beta}
        +\frac{\del P_N^{(\ell)}}{\del Y} \Big|_{\beta},
    \eeq
where the gain and loss terms have now the following structure
    \beq\label{pgb}
    \frac{\del P_N^{(g)}(u_1 ... u_N)}{\del Y} \Big|_{\beta}=
    \sum\limits_{i} \int\limits_{vw}
    \beta(vw|u_i) P_{N+1}(u_1...\slashed{u}_i... u_N
    vw),
    \eeq

    \beq\label{plb}
    \frac{\del P_N^{(l)}(u_1 ... u_N)}{\del Y} \Big|_{\beta}=
    -\sum\limits_{i\neq j} \int\limits_{v}
    \beta(u_i u_j|v) P_N (u_1...u_N),
    \eeq
where $\beta(u_i u_j|v)=\beta(u_j u_i|v)$ is the
(positive--semidefinite) differential probability that the dipoles
$u_i$ and $u_j$ recombine with each other and thus produce the
dipole $v$. The probability conservation law (\ref{probcons}) is
again satisfied, and the analog of Eq.~(\ref{sprobcons}) reads
    \beq\label{sprobconsb}
    \int \dif \Gamma_N
    \frac{\del P_N^{(g)}}{\del Y} \Big|_{\beta} +
    \int \dif \Gamma_{N+1}
    \frac{\del P_{N+1}^{(\ell)}}{\del Y} \Big|_{\beta} =0.
    \eeq
The evolution equation for the generic operator $\cal{O}$ is
obtained as
    \beq\label{Oevolb}
    \frac{\del \lan \cal{O} \ran}{\del Y} \Big|_{\beta} =
    \sum\limits_{N=1}^{\infty}
    \frac{\dif \bar{\Gamma}_{N+2}}{(N\!-\!1)!}\,
    &&\beta(u_N u_{N+1}|u_{N+2})
    P_N(u_1...u_{N+1})
    \nn
    &&\times \big[
    -\cal{O}_{N+1}(u_1...u_{N+1})
    +\cal{O}_N(u_1...u_{N-1}u_{N+2})
    \big].
    \eeq
After straightforward algebra we arrive at the following equation
for the average density
    \beq\label{nevolb}
    \frac{\del \lan n(v) \ran}{\del Y}\Big|_{\beta}=
    \int\limits_{w_1w_2}
    \left[
    -2\beta(w_1v|w_2)\lan n^{(2)}(w_1 v) \ran
    +\beta(w_1w_2|v)\lan n^{(2)}(w_1w_2) \ran
    \right],
    \eeq
and for the average pair density
    \beq\label{n2evolb}
    \frac{\del \lan n^{(2)}(v_1v_2) \ran}{\del Y}\Big|_{\beta}=
    &&\int\limits_{w_1w_2}
    \left[
    -2\beta(w_1v_1|w_2)\lan n^{(3)}(w_1v_1 v_2) \ran
    +\beta(w_1w_2|v_1)\lan n^{(3)}(w_1w_2v_2) \ran
    \right]
    \nn
    &&+\, \big\{v_1 \leftrightarrow v_2 \big\}
    \,- \,2\int\limits_w \beta(v_1v_2|w) \lan n^{(2)}(v_1 v_2) \ran.
    \eeq
As manifest on these equations, the recombination process has two
opposite consequences: on one hand, it yields a negative
contribution to, say, the average number density $\lan n(v) \ran$,
because a dipole $v$ can merge with any other dipole $w_1$  which
is available in the system, and thus disappear; on the other hand,
it yields a positive contribution to $\lan n(v) \ran$, because two
arbitrary dipoles $w_1$ and $w_2$ can fuse with each other and
thus create the dipole $v$. Besides, for higher--point
correlations $\langle n^{(\kappa)} \rangle$ with $\kappa\ge 2$
there are additional negative contributions since any two among
the $\kappa$ external dipoles can recombine with each other.

Yet, it is easy to check that the negative contribution to ${\del
\lan n(v) \ran}/{\del Y}$ prevails after integration over $v$, and
thus the global effect of the recombination is to reduce the total
number of dipoles in the system, as intuitive. Indeed,
Eq.~(\ref{nevolb}) implies:
 \beq\label{intnevolb}
    \int\limits_{v} \frac{\del \lan n(v) \ran}{\del Y}\Big|_{\beta}=
    - \int\limits_{vw_1w_2}
   \beta(w_1w_2|v)\lan n^{(2)}(w_1w_2) \ran\ < \ 0.
    \eeq
(By a similar argument, one can convince oneself that the global
effect of splitting is to increase the total particle number.) In
particular, if the vertex is fully local, $\beta(w_1w_2|v)\propto
\delta(w_1-v)\delta(w_2-v)$, then the effect of the recombination
is to decrease the particle number density at any $v$.

Now let us compare the above equation for $\lan n(v) \ran$ to the
QCD equation (\ref{D1evolR}), as generated by the merging piece
(the second line) in the Hamiltonian (\ref{HamD}).  One then finds
that, {\em formally}, Eq.~(\ref{D1evolR}) can be written in the
form (\ref{nevolb}) by choosing $\beta(w_1w_2|v)=\beta_{3{\mathbb
P}}(\bm{w}_1 \bar{\bm{w}}_1;\bm{w}_2
\bar{\bm{w}}_2|\bm{v}\bar{\bm{v}})$ with
 \beq\label{wrongbeta}
 \beta_{3{\mathbb P}}(\bm{u}_1\bm{v}_1; \bm{u}_2\bm{v}_2|\bm{u}\bm{v})
 \,\equiv\,
 -\,\frac{1}{2}\,\left(\frac{\alpha_s}{2\pi}\right)^2
    \frac{\abar}{2\pi}\, 
    \cal{R}(\bm{u}_1,\bm{v}_1; \bm{u}_2,
    \bm{v}_2|\bm{u},\bm{v})\, + \,
    \big\{1 \leftrightarrow 2 \big\}\,.\eeq
Indeed, with this choice, the first term in Eq.~(\ref{nevolb})
vanishes identically (recall that $\cal{R}$ is a total derivative
w.r.t. its arguments $\bm{u}$ and $\bm{v}$), while the second term
there becomes the same as the r.h.s. of Eq.~(\ref{D1evolR}).
However, it is obvious that such a choice for $\beta$  is
physically meaningless: \texttt{(i)} The vertex (\ref{wrongbeta})
has no fixed sign, as already discussed after Eq.~(\ref{D1evolR}),
and thus cannot be interpreted as a merging rate. \texttt{(ii)}
With this choice, the would--be dominant term in the r.h.s. of
Eq.~(\ref{nevolb}) (namely, the term with an overall factor $-2$)
appears to vanish, and the whole non--trivial contribution comes
from the second term there. \texttt{(iii)} For
$\beta=\beta_{3{\mathbb P}}$, the r.h.s. of Eq.~(\ref{intnevolb})
is exactly zero, showing that the QCD equation (\ref{D1evolR})
cannot be really interpreted as describing dipole merging.

We thus conclude that, when the evolution is viewed as gluon
evolution in the target, the non--linear terms (unitarity
corrections) in the equations for the scattering amplitudes cannot
be attributed to a process like the recombination of dipoles. In
the following section we shall see that a similar problem appears
also when we evolve the projectile, except that, in that case, the
problem arises for the fluctuation terms.

\section{Dipoles in the projectile}
\setcounter{equation}{0}\label{SECT_PROJ}

In this section we shall consider
Eqs.~(\ref{T1evol})--(\ref{T2evol}) for the scattering amplitudes
from the point of view of projectile evolution and show that, in
this context too, the dipole interpretation is plagued with the
difficulty identified before: namely, it necessarily involves the
`dipole recombination' vertex of Eq.~(\ref{wrongbeta}), which has
no probabilistic meaning. We shall discuss two a priori different
strategies to introduce dipoles in the projectile, and show that
they both suffer from the deficiency alluded to above:

First, in Sect. \ref{SECT_HS}, we shall reformulate the
Hamiltonian underlying the QCD equations (cf. Sect.
\ref{SECT_ALPHA}) in terms of new degrees of freedom which are
naturally interpreted as dipoles in the projectile. Then, in Sect.
\ref{SECT_PSPLI} we shall assume --- following Ref. \cite{LL05}
--- that the projectile evolves according to a dipolar
reaction--diffusion process, which involves both splitting and
recombination. The equations for the scattering amplitudes that we
shall obtain in this way will be finally compared to the actual
QCD equations in Sect. \ref{SEC_QCD}.

\subsection{Hamiltonian in the $S$--representation}
\label{SECT_HS}

Returning to the evolution Hamiltonian introduced in Sect.
\ref{SECT_ALPHA}, we shall rewrite it here in a different form ---
the ``$S$--representation'' ---, in which the explicit degrees of
freedom are dipoles in the {\em projectile}, or, more precisely,
the $S$--matrices for the scattering between such dipoles and the
target. Specifically, we introduce the following operators
    \beq\label{Sdag}
    S^{\dagger}_{\bm{x}\bm{y}} = 1 - \frac{g^2}{4N_c}
    \left[
    \alpha^a(\bm{x}) - \alpha^a(\bm{y})
    \right]^2
    \eeq

    \beq\label{S}
    S_{\bm{x}\bm{y}}\, =\,
    \frac{1}{g^2N_c}\,
    \frac{\delta}{\delta \alpha^a(\bm{x})}
    \frac{\delta}{\delta \alpha^a(\bm{y})}
    \eeq
which in the large--$N_c$ limit satisfy
    \beq\label{Scom}
    \left[
    S_{\bm{x}\bm{y}},S^{\dagger}_{\bm{u}\bm{v}}
    \right]=
    \frac{1}{2} \left(
    \delta_{\bm{x}\bm{u}} \delta_{\bm{y}\bm{v}}+
    \delta_{\bm{x}\bm{v}}\delta_{\bm{y}\bm{u}}
    \right),
    \eeq
and in terms of which the Hamiltonian reads
    \beq\label{HamS}
    H\left[S,S^{\dagger}\right] =&&
    \frac{\abar}{2 \pi}
    \int\limits_{\bm{u},\bm{v},\bm{z}}
    \cal{M}(\bm{u},\bm{v},\bm{z})
    \left[
    -S^{\dagger}_{\bm{u}\bm{v}}
    + S^{\dagger}_{\bm{u}\bm{z}} S^{\dagger}_{\bm{z}\bm{v}}
    \right]
    S_{\bm{u}\bm{v}}
    \nn
    &&-\frac{1}{2} \left(\frac{\alpha_s}{2\pi}\right)^2 \atpi
    \int
    \cal{R}(\bm{u}_1,\bm{v}_1; \bm{u}_2, \bm{v}_2|\bm{u},\bm{v})\,
    S^{\dagger}_{\bm{u}\bm{v}}
    S_{\bm{u}_1\bm{v}_1}
    S_{\bm{u}_2\bm{v}_2},
    \eeq
where the expression in the first line is the sum $H_0+H_{2\to
1}$, cf. Eqs.~(\ref{HBFKL}) and (\ref{H21}), while that in the
second line is a rewriting of $H_{1\to 2}$, Eq.~(\ref{MSW}).

Note that the above $S$--form of the Hamiltonian is {\em dual} to
its previous $D$--form, Eq.~(\ref{HamD}), where the {\em duality
transformation} \cite{KL3,BIIT05} consists in replacing
    \beq\label{dual}
    D \rightarrow S \quad,\quad D^{\dag} \rightarrow S^{\dag},
    \eeq
and then exchanging the relative ordering of the
operators\footnote{This permutation corresponds to Hermitian
conjugation, in the sense of an integration by parts in the color
glass functional integral yielding the expectation value over the
target wavefunction (see the discussion in Ref. \cite{BIIT05}).}
$S^{\dagger}$ and $S$. Note also that the splitting and merging
pieces in the original Hamiltonian (\ref{HBFKL})--(\ref{H21}) get
interchanged by this transformation: The merging piece $H_{2\to
1}$, which is the same as the second line of Eq.~(\ref{HamD}), is
dual to the expression in the second line of Eq.~(\ref{HamS}),
which represents the splitting piece $H_{1\to 2}$. This symmetry
reflects the {\em self--duality} of the complete Hamiltonian under
the following transformations \cite{BIIT05}:
 \beq\label{DUAL}
 \alpha^a(\bm{x})\,\longleftrightarrow \,
  i\frac{\delta}{\delta \rho_a(\bm{x})}\,,\qquad
 \frac{1}{i}\, \frac{\delta}{\delta \alpha^a(\bm{x})}\,
 \longleftrightarrow \,
 \rho_a(\bm{x})\,,\eeq
(followed by Hermitian conjugation), a property that has been
actually used in Ref. \cite{BIIT05} to construct the effective
Hamiltonian. As it should become clear from the subsequent
discussion, the physical meaning of the duality transformation is
to exchange dipoles in the target with those in the projectile.

Indeed, the operator $S^{\dagger}_{\bm{x}\bm{y}}$ is recognized as
the $S$--matrix for the scattering between the projectile dipole
$({\bm{x},\bm{y}})$ and the target in the two--gluon exchange
approximation. Since the average amplitude $\lan T\ran$ is
obtained as $\lan T\ran =\langle 1-S^{\dagger}\rangle$, it is in
fact preferable to work in the $S^{\dagger}$--representation, in
which $S^{\dagger}$ is a pure function to be denoted as ${\cal
S}$, and where $T=1-{\cal S}$. This is what we shall properly mean
by the ``$S$--representation'' in what follows. Then, the operator
$S$ can be represented as:
 \beq\label{SSDeltaS}
 S_{\bm{x}\bm{y}} = \frac{\delta}{\delta
 {\cal S}_{\bm{x}\bm{y}}}\,,\qquad \mbox{with}\qquad \frac{\delta}{\delta
 {\cal S}_{\bm{x}\bm{y}}}\,{\cal S}_{\bm{u}\bm{v}}\,\equiv\,
 \frac{1}{2} \left(
    \delta_{\bm{x}\bm{u}} \delta_{\bm{y}\bm{v}}+
    \delta_{\bm{x}\bm{v}}\delta_{\bm{y}\bm{u}}
    \right).
    \eeq
The commutation relation (\ref{Scom}) suggests the use of a
Fock--space language, where $S^{\dagger}$ and $S$ are interpreted
as creation and annihilation operators for dipoles in the
projectile. More precisely, they create and, respectively,
annihilate {\em Pomerons}, where a `Pomeron' refers to the
$S$--matrix for an external dipole. Then, the Hamiltonian
(\ref{HamS}) can be viewed as an {\em effective theory for
Pomerons} \cite{BIIT05}. The first line of Eq.~(\ref{HamS})
describes the BFKL evolution of a Pomeron together with its
dissociation : one Pomeron can split into two via the
`triple--Pomeron {\em splitting} vertex' (\ref{atoM}), which is
essentially the dipole kernel (\ref{dipkernel}).  The second line
in Eq.~(\ref{HamS}) is similarly interpreted as the recombination
of two Pomerons into one, via the `triple--Pomeron {\em merging}
vertex' (\ref{wrongbeta}) \cite{BW93,BV99}. It is easily verified
that the evolution equations for Pomeron correlations generated in
this way, namely
    \beq\label{HSonS}
    \frac{\del \langle {\cal S}^{(\kappa)} \rangle}{\del Y}
    =\lan H {\cal S}^{(\kappa)}\ran,
    \eeq
(where ${\cal S}^{(\kappa)}={\cal S}_1{\cal S}_2\dots {\cal
S}_\kappa$) reproduce indeed the expected equations for the
scattering amplitudes, cf. Eqs.~(\ref{T1evol})--(\ref{T2evol}).
Specifically, the Pomeron splitting piece of the Hamiltonian (the
first line of Eq.~(\ref{HamS})) generates the BFKL terms and the
unitarity corrections, as originally noticed by Janik
\cite{Janik}, whereas the Pomeron merging piece is responsible for
the fluctuation terms \cite{IT04,MSW05,IT05}.

But, clearly, this effective theory cannot be given a
probabilistic interpretation --- as a `reaction--diffusion process
with Pomerons' ---, for the same reason as for its dual
counterpart discussed in Sects. \ref{SECT_HD} and \ref{SECT_TDIP}:
namely, the sign problem of the triple--Pomeron vertex for
merging, Eq.~(\ref{wrongbeta}). One may hope to circumvent this
problem by reformulating the stochastic dynamics directly in terms
of (projectile) dipoles, rather than the associated `Pomerons'.
However, the subsequent analysis will show that this is not the
case.

\subsection{Reaction--diffusion
process in the projectile} \label{SECT_PSPLI}

In this section, we shall study what should be the consequences
--- in terms of evolution equations for dipole scattering amplitudes ---
of a dipolar reaction--diffusion process taking place at the level
of the projectile. That is, we shall assume that the projectile is
a collection of dipoles which evolve via splitting and
recombination with generic vertices, and we shall use the
evolution equations for dipole densities established in Sect.
\ref{SEC_MASTER} to deduce equations for the scattering amplitudes
between these dipoles and a generic target.

In doing so, we shall follow a strategy pioneered by Kovchegov
\cite{K} in his construction of the Balitsky--Kovchegov equation
\cite{B,K}, which is interesting in that it allows one to free
oneself from the two--gluon exchange approximation: the obtained
equations for the scattering amplitudes include multiple
scattering to all orders. This method has been subsequently used
by Levin and Lublinsky to give a rapid derivation of the
large--$N_c$ version of the Balitsky--JIMWLK hierarchy
\cite{LL03,LL04}. In these early applications, the dipole
evolution was restricted to splitting. More recently, Levin and
Lublinsky have attempted to also include dipole recombination
\cite{LL05}, but their final equations are plagued with some
errors of sign that we shall correct in what follows.

Before we proceed with our analysis, let us stress once again that
the evolution to be described below does not represent the actual,
physical, evolution of the projectile in QCD at large--$N_c$ (not
even effectively !), but is only a fictitious reaction--diffusion
process, whose consequences will be compared to the actual QCD
equations in Sect. \ref{SEC_QCD}.

The generic Master equation describing the would--be evolution of
a system of dipoles in the projectile has been already written
down in  Sects. \ref{SEC_MSPLIT} and \ref{SEC_MMERG}. The only
other ingredient that we need for the present purposes is a
factorization formula relating the projectile--target scattering
amplitude to the evolving distribution of dipoles in the
projectile. Following Kovchegov \cite{K}, we write this amplitude
as
    \beq\label{totalt}
    {\cal A}(Y) =
    \sum\limits_{\kappa=1}^{\infty}
    \int \dif \Gamma_{\kappa}\,
    (-1)^{\kappa+1}\,
    n^{(\kappa)}(u_1...u_N;Y-Y_0)\,
    T^{(\kappa)}(u_1...u_N;Y_0),
    \eeq
where $Y_0$ is an arbitrary intermediate rapidity with $0 \le Y_0
\le Y$, $n^{(\kappa)}(Y-Y_0)$ is the $\kappa$--dipole density in a
projectile which has been evolved over $Y-Y_0$ units of rapidity,
and $T^{(\kappa)}(Y_0)$ denotes, as usual, the amplitude for the
scattering between a set of $\kappa$ dipoles and a target evolved
over $Y_0$ units of rapidity. Note that, in writing
Eq.~(\ref{totalt}), no restriction has been assumed on either the
structure of the scattering amplitude for the individual dipoles,
or on the number $\kappa$ of dipoles which scatter simultaneously.
Thus, Eq.~(\ref{totalt}) includes multiple scattering to all
orders, as announced.

Requiring that the total amplitude be insensitive to the rapidity
divider $Y_0$ between the target and the projectile (i.e., to the
choice of a Lorentz frame):
    \beq\label{dift}
    \frac{\dif {\cal A}(Y)}{\dif Y_0} =0,
    \eeq
then using the evolution equations for the dipole densities (cf.
Sect. \ref{SEC_MASTER}) and the fact that these densities are
numerically independent, one eventually obtains an hierarchy for
the scattering amplitudes of projectiles with a given number of
dipoles \cite{LL03,LL04,LL05}.

As before, we shall separately consider the contributions due to
dipole splitting and merging. For the splitting part, the results
of Sect. \ref{SEC_MSPLIT} together with the argument above imply
    \beq\label{T1gen}
    \frac{\del T(x)}{\del Y}\Big|_{\alpha}=
    \int\limits_{uv}
    \alpha(x|uv)
    \left[
    -T(x)+ T(u)+ T(v) - T^{(2)}(uv)
    \right]
    \eeq
and furthermore
    \beq\label{T2gen}
    \frac{\del T^{(2)}(x_1x_2)}{\del Y}\Big|_{\alpha}=
    && \int\limits_{uv}
    \alpha(x_1|uv)
    \left[
    -T^{(2)}(x_1x_2)+ 2 T^{(2)}(vx_2)
    -T^{(3)}(uvx_2)\right]
    \nn
    && +\, \big\{x_1 \leftrightarrow x_2 \big\}.
    \eeq
Higher equations in this hierarchy can be similarly obtained
\cite{LL04}.

The physical interpretation of these equations becomes more
transparent if one recalls that a projectile dipole is more
properly identified with the corresponding $S$--matrix ${\cal S} =
1-T$ (which describes its survival probability), rather than with
its scattering amplitude $T$. Replacing $T$ by $1-{\cal S}$ in
Eq.~(\ref{T1gen}) yields
 \beq\label{S1gen}
    \frac{\del {\cal S}(x)}{\del Y}\Big|_{\alpha}=
    \int\limits_{uv}
    \alpha(x|uv)
    \big[
    -{\cal S}(x)+ {\cal S}^{(2)}(uv)
    \big]
    \eeq
where we have used the fact that $T^{(2)}(uv) =\langle T(u) T(v)
\rangle$ together with the simplified notations ${\cal S}(x)\equiv
\langle {\cal S}(x)\rangle$ and ${\cal S}^{(2)}(uv)\equiv \langle
{\cal S}(u) {\cal S}(v)\rangle$. The evolution described by this
equation is then interpreted as follows : When increasing $Y$ by
$dY$, the original dipole $x$ can either split into a pair of
dipoles $(uv)$ with a probability density $\alpha(x|uv) dY$, or
remain in its original state with a reduced probability
$1-\lambda\, dY$, where $\lambda = \int_{uv} \alpha(x|uv)$ is the
inclusive splitting rate.

Now, let us assume that the dipoles in the projectile can also
recombine, with some generic vertex $\beta$. Then, the evolution
of the scattering amplitudes acquires additional contributions,
which can be inferred from Eqs.~(\ref{nevolb}) and (\ref{n2evolb})
for the dipole densities, together with
Eqs.~(\ref{totalt})--(\ref{dift}). One thus finds:
    \beq\label{T2genb}
    \frac{\del T^{(2)}(x_1x_2)}{\del Y}\Big|_{\beta}=
    && \int\limits_v
    \beta(x_1x_2|v)
    \left[
    T(x_1)+T(x_2)-T(v) - T^{(2)}(x_1x_2) \right].
    \eeq
Once again, the physical interpretation becomes more transparent
after replacing $T=1-{\cal S}$, which yields
 \beq\label{S2genb}
    \frac{\del {\cal S}^{(2)}(x_1x_2)}{\del Y}\Big|_{\beta}=
    && \int\limits_v
    \beta(x_1x_2|v)
    \left[ - {\cal S}^{(2)}(x_1x_2)+ {\cal S}(v) \right].
    \eeq
Namely, the two incoming dipoles $x_1$ and $x_2$ can either
recombine with each other into a dipole $v$ with probability
density $\beta(x_1x_2|v)dY$, or survive in the original
two--dipole configuration with the reduced probability $1 -
dY\int_v \beta(x_1x_2|v)$.

Let us finally notice an error of sign in the corresponding
calculation in Ref. \cite{LL05}: the analog of Eq.~(\ref{T2genb})
--- that is, the contribution of the recombination process to the
evolution of the scattering amplitude
--- appears in Ref. \cite{LL05} with a
sign which is opposite to ours\footnote{See Eq.~(2.19) of Ref.
\cite{LL05}, and focus on the terms involving $\gamma_{n-1}$ (the
analog of our $T^{(\kappa -1)}$) in the r.h.s. of the equation for
$\gamma_n$ (our $T^{(\kappa)}$). The vertex $\Gamma_{2\to 1}$
which appears there corresponds to our $\beta$. Clearly, the terms
proportional to $\gamma_{n-1}$ in the r.h.s. of Eq.~(2.19) have
the opposite sign as compared to our Eq.~(\ref{T2genb}), or the
more general equation (\ref{AEQST}) in the Appendix.}. In view of
the limpid physical interpretation of this equation given above,
it is quite clear that the sign appearing in Ref. \cite{LL05} is
in fact wrong.

\subsection{Comparison with the evolution equations in QCD}
\label{SEC_QCD}

In this section, we shall compare the equations for the scattering
amplitudes derived previously from the Master equation to the
actual evolution equations in QCD, and thus conclude that the
latter cannot be reproduced by a (probabilistically meaningful)
dipole reaction--diffusion process within the projectile.

Of course, there is no a problem at all in so far as the {\em
splitting} alone is concerned: With the splitting vertex $\alpha$
identified with the dipole vertex in Eq.~(\ref{atoM}), the above
equations (\ref{T1gen}) and (\ref{T2gen}) are immediately
recognized as the first two equations of the Balitsky--JIMWLK
hierarchy at large--$N_c$, in agreement with the conclusions in
Refs. \cite{K,LL04}.

Rather, the problem comes again from the {\em recombination}
effects, as encoded e.g. in Eq.~(\ref{T2genb}). When comparing the
terms linear in $T$ in the r.h.s. of this equation to the
fluctuation term in the corresponding QCD equation (\ref{T2evol}),
one comes across the same difficulty as previously discussed
(towards the end of Sect. \ref{SEC_MMERG}) in the context of
target evolution. Namely, in order to reproduce the correct
fluctuation term, one should identify the recombination vertex
$\beta$ with the triple--Pomeron vertex in Eq.~(\ref{wrongbeta}).
With this choice, the first two terms in Eq.~(\ref{T2genb}) would
vanish (since the vertex (\ref{wrongbeta}) is a total derivative
w.r.t. its last argument), while the third term there, which is
surviving, would reproduce the fluctuation term in
Eq.~(\ref{T2evol}). But as repeatedly stressed, the vertex
(\ref{wrongbeta}) has no fixed sign and thus cannot be used within
a physical reaction--diffusion process. The incongruous nature of
this choice becomes even more striking when one realizes that the
would--be surviving term $\propto(-T(v))$ in the r.h.s. of
Eq.~(\ref{T2genb}) yields a {\em negative} contribution in any
physical recombination process (where $\beta$ is positive
semi--definite), whereas the fluctuation term that we need to
reproduce here is rather {\em positive} ! Thus, the fluctuation
terms in the equations for the scattering amplitudes cannot be
reproduced, not even effectively, by a dipole recombination
process in the projectile.

Incidentally, the vertex (\ref{wrongbeta}) which is needed to {\em
formally} relate the prediction (\ref{T2genb}) of the
recombination process to the actual fluctuation term in QCD has
the opposite sign as compared to the corresponding vertex used in
Ref. \cite{LL05} (see Eq.~(3.30) there). The reason why the
correct sign has been nevertheless reported in the final equations
in Ref. \cite{LL05} is because this sign difference in the vertex
has been compensated there by a sign error in the equations for
the scattering amplitudes, as mentioned in the previous
subsection.

Although the overall sign in front of the triple--Pomeron vertex
(\ref{wrongbeta}) is conceptually unimportant --- it does  not
affect our conclusion about the lack of a probabilistic
interpretation ---, it is nevertheless interesting to clarify the
sign difference between our vertex (\ref{wrongbeta}) and that in
Eq.~(3.30) of Ref. \cite{LL05}. To that aim, we shall re-examine
the argument used to obtain this vertex in Ref. \cite{LL05} and
show that, when carefully implemented, this argument leads in fact
to the sign displayed in our Eq.~(\ref{wrongbeta}). Perhaps the
main interest of the subsequent discussion is to demonstrate that,
also in the context of Ref. \cite{LL05}, the `dipole
recombination' vertex has been obtained through a {\em formal
re-interpretation} (similar to our above identification between
the actual fluctuation term in Eq.~(\ref{T2evol}) and the terms
linear in $T$ in Eq.~(\ref{T2genb})), which does not presuppose in
any way the existence of an actual, physical, dipole recombination
process.

\begin{figure}[t]
\begin{center}
    \epsfig{file=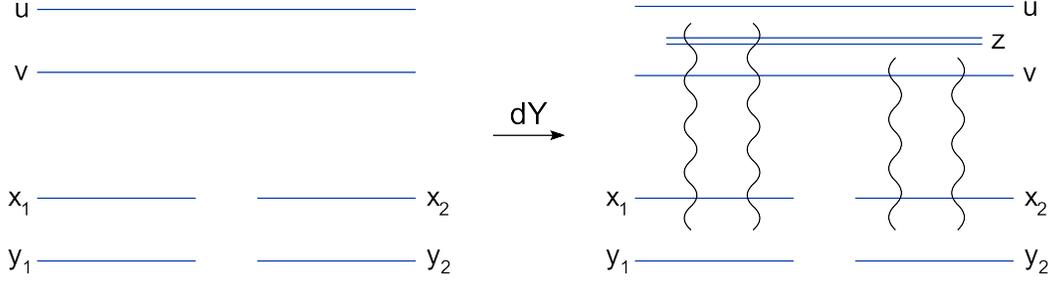,width=14cm}
    \caption{\sl Dipole splitting in the target
    \label{FigVsplit}}\vspace*{0.8cm}
\end{center}
\end{figure}

\begin{figure}[t]
\begin{center}
    \epsfig{file=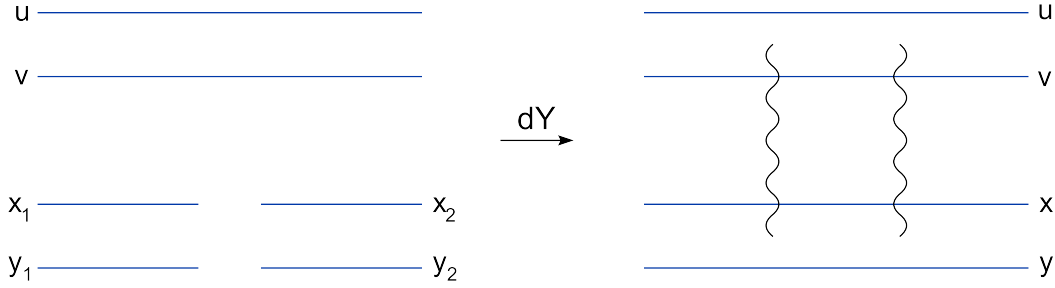,width=14cm}
    \caption{\sl Dipoles (effectively) ``recombining'' in the projectile
    \label{FigVmerge}}\vspace*{1.cm}
\end{center}
\end{figure}

For definiteness, let us consider the fluctuation terms in the
equations for the scattering amplitudes. (By exchanging the roles
of the projectile and the target, the following argument can be
easily translated to the unitarity corrections.) As we have seen,
these terms have indeed a dipole interpretation at large $N_c$,
namely, they correspond to dipole {\em splitting} in the {\em
target}. Loosely speaking, splitting in the target corresponds to
merging in the projectile (as this amounts to viewing the same
Feynman diagrams upside down !), which makes it tempting to try
and interpret the fluctuation terms as the result of dipole {\em
merging} in the {\em projectile}. This in turn amounts to
identifying the actual QCD process described in
Fig.~\ref{FigVsplit} to the fictitious process illustrated in
Fig.~\ref{FigVmerge}, and then use this matching condition to
extract an effective `dipole recombination' vertex. As we shall
shortly see, this procedure yields indeed the vertex shown in
Eq.~(\ref{wrongbeta}), including the sign.

Specifically, the diagram on the r.h.s. of Fig.~\ref{FigVsplit} is
computed as
    \beq\label{Vsplit}
    V_1 = &&\atpi \int\limits_{\bm{z}}
    \mathcal{M}(\bm{u},\bm{v},\bm{z})
    \left[-\alpha_s^2 \cal{A}_0(\bm{x}_1,\bm{y}_1|\bm{u},\bm{z}) \right]
    \left[-\alpha_s^2 \cal{A}_0(\bm{x}_2,\bm{y}_2|\bm{z},\bm{v}) \right]
    + \{1 \rightarrow 2\}.
    \eeq
It is important to notice at this point that the elementary
dipole--dipole scattering is ${\rm i}^2 \alpha_s^2 \cal{A}_0 = -
\alpha_s^2 \cal{A}_0 < 0$, where the two factors of ${\rm i}$ come
from the two propagators of the exchanged gluons. By assumption,
the diagram in the r.h.s. of Fig.~\ref{FigVmerge} is equal to
    \beq\label{Vmerge}
    V_2 = -\alpha_s^2
    \int\limits_{\bm{x}\bm{y}}
    \beta(\bm{x}_1,\bm{y}_1; \bm{x}_2, \bm{y}_2|\bm{x},\bm{y})\,
    \cal{A}_0(\bm{x},\bm{y}|\bm{u},\bm{v}).
    \eeq
Setting $V_1=V_2$, then acting with $\lap{u}\lap{v}$, where
$\bm{u} \neq \bm{v}$, on both sides of the equation and using
$\lap{x} \ln (\bm{x}^2) = 4\pi \delta(\bm{x})$, we finally obtain
$\beta=\beta_{3{\mathbb P}}$, as anticipated. Note that the
difference in sign as compared to the corresponding calculation in
Ref. \cite{LL05} comes from the minus sign in the structure of the
dipole--dipole elementary amplitude (this sign has been overlooked
in Ref. \cite{LL05}). But even though it generates, by
construction, the correct fluctuation terms, the `recombination
vertex' of Eq.~(\ref{wrongbeta}) cannot be used within a
meaningful stochastic process, because of the sign problem
explained before.

Ultimately, the lack of a probabilistic interpretation in terms of
dipole recombination reflects the profound physical difference
between the actual QCD mechanism for gluon saturation --- the
coherent suppression of gluon emission by strong color fields
\cite{SAT,PATH} --- and the `mechanical' $2\to 1$ dipole merging.

\section*{Acknowledgments}
\vspace*{-0.5cm}

We are grateful to Al Mueller for continuous discussions on this
and related subjects, and for a critical reading of the
manuscript. We would like to thanks Larry McLerran, Arif Shoshi
and Anna Sta\'sto for useful comments on the manuscript. G.S. is
funded by the National Funds for Scientific Research (Belgium).

\appendix
\section{Evolution equations for a generic reaction--diffusion process}

In this section we present, for completeness, the general
evolution equations for a generic reaction--diffusion process with
non--local vertices for splitting and merging. Particular cases of
these equations have already appeared in the main body of the
paper.

We consider the reaction--diffusion process $A \rightleftharpoons
A+A$ in which one particle at position $x$ can split into two
particles at positions $x_1$ and $x_2$ at a rate
$\alpha(x|x_1x_2)$, and two particles at positions $x_1$ and $x_2$
can merge into one particle at position $x$ at a rate
$\beta(x_1x_2|x)$. We denote by $P_N(x_1\dots x_N)\equiv P_N(X)$
the probability density to observe a system of $N$ particles at
positions $x_1,\dots,x_N$ at a given time $t$ (the dependence of
$P_N$ upon $t$ is kept implicit). Due to merging and splitting,
the probabilities evolve according to the following hierarchy of
equations, which is usually referred to as the {\em Master
equation} :
\begin{eqnarray}
\partial_t P_N(X)
 &\!=\!& \sum_{i\neq j} \int_z \alpha(z|x_ix_j)
             P_{N-1}(x_1\dots\slashed{x}_i\dots\slashed{x}_j\dots x_Nz)
           - \sum_i\int_{z_1,z_2}\alpha(x_i|z_1z_2) P_N(X)\nonumber\\
 &\!+\!& \sum_i\int_{z_1,z_2}\beta(z_1z_2|x_i) P_{N+1}(x_1\dots\slashed{x}_i\dots x_Nz_1z_2)
           - \sum_{i\neq j} \int_z \beta(x_ix_j|z) P_N(X).
\end{eqnarray}
In writing this equation, we have also assumed the symmetry
properties $\alpha(x|x_1x_2)= \alpha(x|x_2x_1)$ and
$\beta(x_1x_2|x)=\beta(x_2x_1|x)$

Let us now derive the equation for the evolution of an observable
$\cO$ with an average value defined by
\[
\avg{\cO} = \sum_{N=1}^\infty \int d\Gamma_N P_N(x_1\dots x_N)
\cO_N(x_1 \dots x_N),
\]
where in the integral over the phase--space one needs to count
only once the configurations differing only by the order of their
elements (cf. Eq.~(\ref{dgamma})). After some straightforward
algebra, we obtain the following evolution equation for the
observable
\begin{eqnarray}
\partial_t \avg{\cO}
 = \sum_{N=1}^\infty \frac{1}{N!} \int_{x_i} &\!\!\!\!& P_N(X)
  \left\{ \sum_{i=1}^N \int_{z_1,z_2} \alpha(x_i|z_1z_2)
 \left[\cO_{N+1}(x_1\dots\slashed{x}_i\dots x_Nz_1z_2)-\cO_N(X) \right]\right.\nonumber\\
 &\!\!\!\!& +\left. \sum_{i\neq j} \int_z \beta(x_ix_j|z)
 \left[\cO_{N-1}(x_1\dots\slashed{x}_i\dots\slashed{x}_j\dots x_Nz)-
 \cO_N(X)\right]\right\},
\label{eq:avgevol}
\end{eqnarray}
which encompasses Eqs.~(\ref{Oevol}) and (\ref{Oevolb}) of the
main text.

In particular, let us derive the evolution equations for the
average $k$--particle densities. The average particle number
density $\avg{n(y)}$ is defined as in Eq.~(\ref{nave}), and
involves the following density operator
\begin{equation}\label{eq:defn}
n_N(y|x_1\dots x_N) = \sum_{i=1}^N \delta_{yx_i}.
\end{equation}
Similarly, the $k$--particle density operator is defined as
\[
n^{(k)}_N(y_1\dots y_k|x_1\dots x_N)
 = \sum_{i_1\neq\dots\neq i_k} \delta_{y_1x_{i_1}}\dots\delta_{y_kx_{i_k}}.
\]
Inserting the definition \eqref{eq:defn} into the evolution
equation \eqref{eq:avgevol} gives
\begin{eqnarray}
\partial_t\avg{n(y)}
&\!=\!& \int_{z_1,z_2}\alpha(z_1|yz_2)\avg{n(z_1)}
                   +\alpha(z_2|z_1y)\avg{n(z_2)}
                   -\alpha(y|z_1z_2)\avg{n(y)}\label{eq:nevol}\\
&\!\!&+\beta(z_1z_2|y)\avg{n^{(2)}(z_1z_2)}
      -\beta(z_1y|z_2)\avg{n^{(2)}(z_1,y)}
      -\beta(yz_2|z_1)\avg{n^{(2)}(yz_2)}.\nonumber
\end{eqnarray}
In a similar way, but after some lengthier manipulations, one ends
up with the following evolution equation for the general
$k$--particle density:
\begin{eqnarray}
  \lefteqn{\partial_t\avg{n^{(k)}(y_1\dots vy_k)}}\\
& = & \int_{z_1,z_2}\sum_{i=1}^k
      \alpha(z_1|yz_2)\avg{n^{(k)}(y_1\dots y_{i-1}z_1y_{i+1}\dots y_k)}
     -\beta(z_1y|z_2) \avg{n^{(k+1)}(y_1\dots y_kz_1)}\nonumber\\
&   & \phantom{\int_{z_1}\sum}
     +\alpha(z_2|z_1y)\avg{n^{(k)}(y_1\dots y_{i-1}z_2y_{i+1}\dots y_k)}
     -\beta(yz_2|z_1) \avg{n^{(k+1)}(y_1\dots y_kz_2)}\nonumber\\
&   & \phantom{\int_{z_1}\sum}
     -\alpha(y|z_1z_2)\avg{n^{(k)}(y_1\dots y_k)}
     +\beta(z_1z_2|y)\avg{n^{(k+1)}(y_1\dots\slashed{y}_i\dots y_kz_1z_2)}\nonumber\\
& + & \int_z \sum_{i\neq j}
      \alpha(z|y_iy_j)\avg{n^{(k-1)}(y_1\dots\slashed{y}_i\dots\slashed{y}_j\dots y_kz)}
     -\beta(y_iy_j|z)\avg{n^{(k)}(y_1\dots y_k)}.\nonumber
\end{eqnarray}

Finally, we use the density operators to construct a {\em
scattering amplitude} which includes multiple scattering to all
orders. This is defined as \cite{K}
 \beq {\cal A}(t) = \sum_k
 \frac{1}{k \,!}\int_{x_i} (-)^{k+1}
              T^{(k)}(x_1\dots x_k;t_0) \,n^{(k)}(x_1\dots x_k;t-t_0)
 \eeq
with $T^{(k)}$ being the scattering amplitude for a set of $k$
particles. In this expression $t_0$ represents an arbitrary
intermediate time, but the physical quantity ${\cal A}(t)$ is of
course independent of $t_0$ : $d{\cal A}(t)/dt_0 =0$. From this
condition together with the evolution equation for the average
densities derived before, one can finally deduce a set of
evolution equations for the scattering amplitudes. The general
equation in this hierarchy reads:
\begin{eqnarray}\label{AEQST}
\partial_t T^{(k)}(x_1\dots x_k;t)
  &\!=\!& \sum_{i=1}^k \int_{z_1,z_2} \alpha(x_i|z_1z_2)
              \left\lbrack T^{(k)}(x_1\dots\slashed{x}_i\dots x_kz_1;t)
                         + T^{(k)}(x_1\dots\slashed{x}_i\dots x_kz_2;t) \right.
  \nonumber\\[-4mm]
  &\!\! & \phantom{\sum_{i=1}^k z1 z2 \alpha(x_i|z_1z_2)}
              \left.     - T^{(k)}(x_1\dots x_k;t)
                         - T^{(k+1)}(x_1\dots\slashed{x}_i\dots y_kz_1z_2;t) \right\rbrack
  \nonumber\\[-3mm]
  &\!+\!& \sum_{i\neq j} \int_{z_1} \beta(x_ix_j|z)
              \left\lbrack T^{(k-1)}(x_1\dots\slashed{x}_i\dots x_k;t)
                         + T^{(k-1)}(x_1\dots\slashed{x}_j\dots x_k;t) \right.
  \nonumber\\[-4mm]
  &\!\! & \phantom{\sum_{i=1}^k z_1 \alpha(x_i|z_1z_2)}
              \left.     - T^{(k-1)}(x_1\dots\slashed{x}_i\dots\slashed{x}_j\dots y_kz;t)
                         - T^{(k)}(x_1\dots x_k;t) \right\rbrack.
                        \nonumber\\
\end{eqnarray}


\begin{thebibliography}{10}

\bibitem{IM032}
E.~Iancu and A.H.~Mueller, {\it Nucl.\ Phys.}\ {\bf A730} (2004)
494.

\bibitem{MS04}
A.H.~Mueller and A.I.~Shoshi, {\it Nucl.\ Phys.}\ {\bf B692}
(2004) 175.

\bibitem{IMM04}
E.~Iancu, A.H.~Mueller and S.~Munier, {\it Phys.~Lett.~}{\bf B606}
(2005) 342.

\bibitem{IT04}
E.~Iancu and D.N.~Triantafyllopoulos, {\it Nucl.~Phys.~}{\bf A756}
(2005) 419.

\bibitem{B}
I.~Balitsky, {\it Nucl.\ Phys.}\ {\bf B463} (1996) 99; {\it Phys.
Lett.} {\bf
  B518} (2001) 235; {\it ``High-energy QCD and Wilson lines''},
  arXiv:hep-ph/0101042.

\bibitem{JKLW}
J.~Jalilian-Marian, A.~Kovner, A.~Leonidov and H.~Weigert, {\it
Nucl.\ Phys.}\
  {\bf B504} (1997) 415; {\it Phys.\ Rev.}\ {\bf D59} (1999) 014014;
  J.~Jalilian-Marian, A.~Kovner and H.~Weigert, {\it Phys.\ Rev.}\ {\bf D59}
  (1999) 014015; A. Kovner, J. G. Milhano and H. Weigert, {\it Phys. Rev.} {\bf
  D62} (2000) 114005.

\bibitem{RGE}
E.~Iancu, A.~Leonidov and L.~McLerran, {\it Nucl. Phys.}~{\bf
A692} (2001) 583;
  {\it Phys. Lett.} {\bf B510} (2001) 133; E.~Ferreiro, E.~Iancu, A.~Leonidov
  and L.~McLerran, {\it Nucl. Phys.} {\bf A703} (2002) 489.

\bibitem{W}
H.~Weigert, {\it Nucl. Phys.} {\bf A703} (2002) 823.

\bibitem{IT05}
E.~Iancu, D.N.~Triantafyllopoulos, {\it Phys.~Lett.~}{\bf B610}
(2005) 253.

\bibitem{MSW05}
A.H.~Mueller, A.I.~Shoshi, S.M.H.~Wong, {\it Nucl.~Phys.~}{\bf
B715} (2005)
  440.

\bibitem{BIIT05}
J.-P.~Blaizot, E.~Iancu, K.~Itakura, D.N.~Triantafyllopoulos, {\it
  Phys.~Lett.~}{\bf B615} (2005) 221.

\bibitem{LL05}
E.~Levin, M.~Lublinsky, {\it ``Towards a symmetric approach to
high energy
  evolution: generating functional with Pomeron loops''}, arXiv:hep-ph/0501173.

\bibitem{Levin05}
E.~Levin, {\it ``High energy amplitude in the dipole approach with
Pomeron
  loops: asymptotic solution''}, arXiv:hep-ph/0502243.

\bibitem{KL05}
A.~Kovner and M.~Lublinsky, {\it Phys.~Rev.~}{\bf D71} (2005)
085004.

\bibitem{KL3}
A.~Kovner and M.~Lublinsky, {\it Phys.~Rev.~Lett.~}{\bf 94} (2005)
181603.

\bibitem{KL4}
A.~Kovner and M.~Lublinsky, {\it ``Dense-Dilute Duality at work:
dipoles of the
  target''}, arXiv:hep-ph/0503155.

\bibitem{BREM}
Y. Hatta, E. Iancu, L. McLerran, A. Stasto,
D.N.~Triantafyllopoulos, {\it
  ``Effective Hamiltonian for QCD Evolution at High Energy''},
  hep-ph/0504182, to appear in {\it Nucl.Phys.}\ {\bf A}.

\bibitem{MMSW05}
C. Marquet, A.H. Mueller, A.I. Shoshi, and S.M.H. Wong {\it ``On
the
  Projectile-Target Duality of the Color Glass Condensate in the Dipole
  Picture''}, arXiv:hep-ph/0505229.

\bibitem{HIMS05}
Y. Hatta, E. Iancu, L. McLerran, and A. Stasto, {\it `` Color
dipoles from
  Bremsstrahlung in QCD evolution at high energy''},
  arXiv:hep-ph/0505235, to appear in {\it Nucl.\ Phys.}\ {\bf A}.


\bibitem{Balit05}
I. Balitsky, {\it ``High-enegy effective action from scattering of
QCD shock
  waves''}, arXiv:hep-ph/0507237.

\bibitem{KL5}
A.~Kovner and M.~Lublinsky, {\it ``More Remarks on High Energy
Evolution''}, arXiv:hep-ph/0510047.

\bibitem{AM94}
A.H.~Mueller, {\it Nucl. Phys.} {\bf B415} (1994) 373; A.H.
Mueller, B. Patel,
  {\it Nucl. Phys.} {\bf B425} (1994) 471.

\bibitem{AM95}
A.H.~Mueller, {\it Nucl. Phys.} {\bf B437} (1995) 107.

\bibitem{IM031}
E.~Iancu and A.H.~Mueller, {\it Nucl.\ Phys.}\ {\bf A730} (2004)
460.

\bibitem{BFKL}
L.N.~Lipatov, {\it Sov.\ J.\ Nucl.\ Phys.}\,{\bf 23} (1976) 338;\\
E.A.~Kuraev,
  L.N.~Lipatov and V.S.~Fadin, {\it Zh. Eksp. Teor. Fiz} {\bf 72}, 3 (1977)
  ({\it Sov. Phys. JETP }{\bf 45} (1977) 199);\\ Ya.Ya.~Balitsky and
  L.N.~Lipatov, {\it Sov.\ J.\ Nucl.\ Phys.} {\bf 28} (1978) 822.

\bibitem{Salam95}
G.P.~Salam, {\it Nucl. Phys.} {\bf B449} (1995) 589; {\it Nucl.
Phys.} {\bf
  B461} (1996) 512; A.H.~Mueller, G.P.~Salam, {\it Nucl. Phys.} {\bf B475}
  (1996) 293.

\bibitem{K}
Yu.V.~Kovchegov, {\it Phys. Rev.} {\bf D60} (1999), 034008; {\it
ibid.} {\bf
  D61} (1999), 074018.

\bibitem{LL04}
E.~Levin and M.~Lublinsky, {\it Phys. Lett.} {\bf B607} (2005)
131.



\bibitem{MV}
L.~McLerran and R.~Venugopalan, {\it Phys.\ Rev.}\ {\bf D49}
(1994) 2233; {\it
  ibid.} {\bf 49} (1994) 3352; {\it ibid.} {\bf 50} (1994) 2225.

\bibitem{CGCreviews}
  E.~Iancu, A.~Leonidov and L.~McLerran, {\it ``The Colour Glass Condensate: An
  Introduction''}, arXiv:hep-ph/0202270. Published in {\it QCD Perspectives on
  Hot and Dense Matter}, Eds. J.-P.~Blaizot and E.~Iancu, NATO Science Series,
  Kluwer, 2002;\\ E.~Iancu and R.~Venugopalan, {\it ``The Color Glass
  Condensate and High Energy Scattering in QCD''}, arXiv:hep-ph/0303204.
  Published in {\it Quark-Gluon Plasma 3}, Eds. R.C.~Hwa and X.-N.~Wang, World
  Scientific, 2003;\\ H.~Weigert, {\it ``Evolution at small $x_{\rm bj}$: The
  Color Glass Condensate''}, arXiv:hep-ph/0501087.

\bibitem{PATH}
J.-P.~Blaizot, E.~Iancu and H.~Weigert, {\it Nucl.~Phys.~}{\bf
A713} (2003)
  441.

\bibitem{RW03}
K.~Rummukainen and H.~Weigert, {\it Nucl.\ Phys.}\ {\bf A739}
(2004) 183.

\bibitem{BW93}
J.~Bartels and M.~W\"{u}sthoff, {\it Z.~Phys.~}{\bf C66} (1995)
157.

\bibitem{BV99}
M.~Braun and G.P.~Vacca, {\it Eur.~Phys.~J.~}{\bf C6} (1999) 147;
M.~Braun,
  {\it Phys.~Lett.~}{\bf B483} (2000) 115.

\bibitem{Janik}
R.A.~Janik, {\it Phys. Lett.} {\bf B604} (2004) 192.

\bibitem{Gardiner}
C.W. Gardiner, {\it Handbook of Stochastic Methods},  Springer,
Berlin, 2004.

\bibitem{SaarPanja}
W.~Van Saarloos, {\it Phys. Rep.} {\bf 386} (2003) 29; D.~Panja,
Phys. Rep.
  {\bf 393} (2004) 87.

\bibitem{MP03}
S.~Munier and R.~Peschanski, {\it Phys. Rev. Lett.} {\bf 91}
(2003) 232001;
  {\it Phys.\ Rev.}\ {\bf D69} (2004) 034008; {\it ibid.} {\bf D70} (2004)
  077503.

\bibitem{geometric}
A.M.~Stasto, K.~Golec-Biernat and J.~Kwiecinski, Phys. Rev. Lett.
{\bf 86}
  (2001) 596.

\bibitem{SCALING}
E.~Iancu, K.~Itakura, and L.~McLerran, {\it Nucl. Phys.} {\bf
A708} (2002) 327.

\bibitem{MT02}
A.H.~Mueller and D.N.~Triantafyllopoulos, {\it Nucl. Phys.} {\bf
B640} (2002)
  331.

\bibitem{DT02}
D.N.~Triantafyllopoulos, {\it Nucl. Phys.} {\bf B648} (2003) 293.

\bibitem{Kov05}
Yu.V. Kovchegov, {\it ``Inclusive Gluon Production In High Energy
Onium-Onium
  Scattering''}, arXiv:hep-ph/0508276.

\bibitem{SAT}
E.~Iancu and L.~McLerran, {\it Phys. Lett.} {\bf B510} (2001) 145.

\bibitem{AM02}
A. H. Mueller,  {\it Nucl. Phys.} {\bf B643} (2002) 501.


\bibitem{LL03}
E.~Levin and M.~Lublinsky, {\it Nucl.\ Phys.}\ {\bf A730} (2004)
191.

\bibitem{GS05}
G. Soyez, {\it Phys. Rev.} {\bf D72} (2005) 016007.

\bibitem{EGBM05}
R. Enberg, K. Golec--Biernat, S. Munier, {\it ``The high energy
asymptotics of
  scattering processes in QCD''}, arXiv:hep-ph/0505101.

\bibitem{ODDERON}
Y.~Hatta, E.~Iancu, K.~Itakura and L.~McLerran, {\it Nucl. Phys.}
{\bf A760}
  (2005) 172.

\end{thebibliography}

\end{document}